\documentstyle[12pt]{article}
\renewcommand{\theequation}{\arabic{section}.\arabic{equation}}
\renewcommand{\thesection}{ \setcounter{equation}{0} \Roman{section}.}

\renewcommand{\appendix}
 { \setcounter{section}{0}  \setcounter{subsection}{0}
 \renewcommand{\thesection}{ \setcounter{equation}{0}
 Appendix \Alph{section}:}
 \renewcommand{\theequation}{\Alph{section}.\arabic{equation}}}
\begin{document}
\vskip 2.cm
\
\begin{center}
\vskip 2.cm
\
{\Large\bf Consistent  low-energy reduction of the three-band model
for copper oxides with O-O hopping to the effective $t$-$J$ model \\}
\vskip 2.cm
{\large V.I. Belinicher and  A.L. Chernyshev }\\
Institute of Semiconductor Physics, 630090, Novosibirsk, Russia \\
\vskip 2.cm
\end{center}

\begin{abstract}
A full three-band model for the CuO$_{2}$ plane with inclusion of all
essential interactions - Cu-O and O-O hopping, repulsion at the
copper and oxygen and between them - is considered.
A general procedure of the low-energy reduction of the primary
Hamiltonian to the Hamiltonian of the generalized $t$-$t'$-$J$ model
is developed.  An important role of the direct O-O hopping is
discussed.   Parameters of the effective low-energy model
(the hopping integral, the band position and the superexchange
constant $J$ are calculated.
An analysis of the obtained  data shows that  the experimental value
of $J$ fixes the charge transfer energy $\Delta =(\epsilon
_{p}-\epsilon _{d})$  in a narrow region of energies.
\end{abstract}
\newpage
\section{Introduction}\ \ \
Since the discovery of high-T$_{c}$ superconductors, major
theoretical efforts were devoted to finding the simplest model which  would
contain all details relevant to superconductivity.
Anderson first suggested \cite{An} that the two-dimensional
single-band $t$-$J$ model fits well for this role.

Now there is a general agreement that the CuO$_{2}$ planes are common to
all high-temperature superconductors. A realistic model for these
planes was proposed from the first principles by Emery \cite{Em1},
Varma, Schmitt - Rink and Abrahams \cite{Va}, and Gaididei and Loktev
\cite{Lok}.

During last years there was a polemic about the equivalency of the $t$-$J$
model and the Emery model \cite{Em1} in the low-energy limit. The
principal step was made in the work of Zhang and Rice \cite{Zh1}, where
they proposed the idea of the local singlet and idea of the Wannier
representation for O-states. The problem of the low-energy
reduction has been intensively investigated \cite{Zh2}-\cite{Yu1}.
The singlet - triplet effective Hamiltonian was obtained in works of
Yushankhai and Lovtsov \cite{Yu1} and Shen and Ting \cite{Sh1}.  In our
recent work \cite{Be1} the quantitative comparison of the exact solution for
the three-band model with the solution for the generalized $t$-$J$ model
was performed. However, the situation seems not completely clarified
since (1) most of the above-named works were dealing
with the unrealistic region of parameters $t \ll U_{d}, \Delta $,
 where $t$ is the Cu-O hopping integral, $U_{d}$ is the Coulomb repulsion
at the Cu site, $\Delta=(\epsilon _{p} -\epsilon_{d})$
 is the charge-transfer energy, and (2) all above named works did not
  consistently take into account the direct O-O hopping.

In this work we develop a general approach to the consistent
low-energy reduction of the three-band model.  For this we following by
Zhang and Rice \cite{Zh1} transform the primary Hamiltonian of the model
from the terms of the usual oxygen
states to the terms of symmetrical and antisymmetrical
orthonormalized oxygen states on O - clusters.  Further we introduce a new
basis of the local states with the certain number of
particles.  Since only  the systems with filling close to unity are of
interest , we restrict ourselves to one- and two-hole local states.
After that, the diagonalization of the local part of the Hamiltonian
presents no special problems. It should  be noted that we consider
the direct O-O hopping.  We have found that it does not change
crucially the picture of the local states  but play an important role
for the effective hopping parameters.  Next, we hold only the lowest
two-hole local singlet state and consider its transitions over the background
of the lowest one-hole (spin) states.  Thus, we get the $t$-$t'$-$J$
model.  All other states we take into account perturbatively, by
applying the canonical transformation.  In this way we obtain the general
form of the second-order corrections to the $t$-$t'$-$J$ model and  the
expression for the superexchange constant $J$ .

At the beginning, we investigate three limiting cases of the complete
three-band model.  Because the Gilbert spaces in these limiting cases
are restricted, mathematical treatment is simplified.
Further, we investigate the general case of the three-band model in the
region of parameters where the charge-transfer insulator \cite{Za1}  is
the ground state of the undoped system.
We conclude that the $t$-$J$ model is valid
for the doped charge-transfer insulator,
 i.e. corrections to it are small. Corrections to the hopping integrals at
the second and other neighbors are not relatively small, and thus the
simple $t$-$t'$-$J$ model does not follow from the three-band model.
Since the experimental value of the superexchange constant $J$ is
known very well, we determine the
value of the charge-transfer energy  $\Delta$
through $J$.  We have found that a narrow region of energies for
$\Delta$ is suitable.

We also consider the repulsion at the O site U$_{p}$ and the repulsion
between the Cu and O ions V$_{pd}$ terms in the framework of
our approach.  Their influence on the values of $\Delta$   and the
first hopping integral $t_{1}$ is discussed.   Thus, we perform
a consistent and full consideration of the three-band model.

In Sec. II we represent the three-band Hamiltonian in terms of
new oxygen states.  In Sec. III  we diagonalize the local Hamiltonian
and rewrite it in terms of the Hubbard operators.  In Sec. IV the
low-energy reduction procedure is developed. In Sec. V three
limiting cases of the three-band model are considered.  In Sec. VI
the quantitative analysis is performed.  In Sec. VII the  U$_{p}$ and
V$_{pd}$ terms are considered.  Section VIII presents our conclusions.
 In Appendices A and B the explicit form of the required matrix
elements are presented.
 In Appendix C the three-hole states are considered.
 In Appendix D the shifts of the energies of local states connected
with U$_{p}$ and V$_{pd}$ terms are presented.

\section{Three-band Hamiltonian}\ \ \
 The three-band model studied in this paper was originally proposed
by Emery \cite{Em1}.  It is given by the following Hamiltonian
\begin{eqnarray}
\label{1}
H&&=\epsilon_{d}\sum_{l,\alpha} n^{d}_{l\alpha}+
\epsilon_{p}\sum_{m,\alpha}n^{p}_{m\alpha}
+U_{d}\sum_{l}n^{d}_{l\uparrow}n^{d}_{l\downarrow}
+\Delta H + H'  \\
\label{1x}
\Delta H \ = \ &&U_{p}\sum_{m} n^{p}_{m\uparrow}n^{p}_{m\downarrow}
+V_{pd}\sum_{\langle lm\rangle,\alpha ,\beta }
n^{d}_{l\alpha }n^{p}_{m \beta },
\end{eqnarray}
where $l$ and $m$ denote summation over the Cu- and O-sites respectively,
$n_{l\alpha}^{d}= d_{l\alpha }^{+}d_{l\alpha }$,
 $n^{p}_{m\alpha }=p_{l\alpha}^{+}p_{l\alpha}$, $ d_{l\alpha }^{+} \
(d_{l\alpha})$
 creates (annihilates) the  Cu (3d$_{x^{2}-y^{2}}$) hole, $p_{l\alpha}^{+}
\ (p_{l\alpha })$  creates (annihilates) an O (p$_{x}$, p$_{y}$)
hole.
U$_{d}$ and U$_{p}$  are intrasite Coulomb repulsion at the copper
and oxygen respectively, V$_{pd}$ is the intersite repulsion between
the nearest Cu and O holes.
The hybridization Hamiltonian $H'$ includes Cu-O and O-O hopping
terms:
\begin{eqnarray}
\label{2}
H'&&=t\sum_{\langle lm\rangle\alpha}(d^{+}_{l\alpha}p_{m\alpha}+ H.c.)
\nonumber\\
&&-t_{p}\sum_{\langle mm'\rangle\alpha}
(p^{+}_{m\alpha}p_{m'\alpha}+ H.c.) \ ,
\end{eqnarray}
where $\langle l,m\rangle$ denotes the
nearest-neighbor Cu and O sites and $\langle mm'\rangle$ denotes the
nearest-neighbor O sites.
The quantities $t$ and $t_{p}$ are positive.  In Eq. (\ref{2}) the sign
convention corresponds to the change of the signs of the operators at the
odd sites, which corresponds to the shift of quasimomentum space by
$(\pi /a,\pi /a)$.  As was proposed in \cite{Fr1}, such transformation makes
all O-O hopping constants of the same sign and negative.

The three-band model (\ref{1}),(\ref{2}) and its modifications have
been studied recently by many authors \cite{Zh2,Em2,ZB}.
 One of the simplifications used in all analytical works was
neglecting  the repulsion at the oxygen and between the nearest - neighbor
copper and oxygen.  To take into account the main effect, we will firstly
consider the model (\ref{1}) - (\ref{2})  without U$_{p}$ and V$_{pd}$ terms
and later include these terms as the perturbation.

The Fourier - transformed Hamiltonian (\ref{1}) - (\ref{2}) with the above
named restriction is given by
\begin{eqnarray}
\label{3}
H_{0}&&=\epsilon_{d}\sum_{{\bf k},\alpha}
d^{+}_{{\bf k}\alpha}d_{{\bf k}\alpha}+
\epsilon_{p}\sum_{{\bf k},\alpha}
(p^{+}_{{\bf k}\alpha x} p_{{\bf k}\alpha x}
+p^{+}_{{\bf k}\alpha y}p_{{\bf k}\alpha y})
\nonumber\\
&&+U_{d}\sum_{\bf k_{1}+k_{2}=k_{3}+k_{4}}
d^{+}_{{\bf k_{1}}\uparrow}d_{{\bf k_{2}}\uparrow}
d^{+}_{{\bf k_{3}}\downarrow}d_{{\bf k_{4}}\downarrow} \ ,
\end{eqnarray}
and
\begin{eqnarray}
\label{4}
H'&&=2t\sum_{{\bf k}\alpha} \{ d^{+}_{{\bf k}\alpha}[p_{{\bf k}\alpha x}
\cos(k_{x}/2)+ p_{{\bf k}\alpha y}\cos(k_{y}/2)]+ H.c.\}
\nonumber\\
&&-t_{p}\sum_{{\bf k}\alpha} \{ 2[\cos((k_{x}+k_{y})/2)
+ \cos((k_{x}-k_{y})/2)]  p^{+}_{{\bf k}\alpha x} p_{{\bf k}\alpha y}
+ H.c.\}
\end{eqnarray}
where
\begin{eqnarray}
\label{4a}
d_{{\bf k} }=\sum_{l} d_{l} \exp(-i{\bf k}{\bf r}_{l}) \ ,
\ p_{{\bf k} x(y)}=\sum_{m \in x(y)} p_{m} \exp(-i{\bf k}{\bf r}_{m}) \ ,
\end{eqnarray}
$m \in \{x(y)\}$   denotes the $x(y)$ O-sublattices.
Hereafter, the lattice constant is equal to unity.  It is reasonable to
introduce the symmetrical and antisymmetrical orthonormalized  $p$-operators
 combination as
\begin{eqnarray}
\label{5}
q_{{\bf k}}&&\ =\ (\cos(k_{x}/2) p_{{\bf k}x}
+\cos(k_{y}/2) p_{{\bf k}y})(1+\gamma_{\bf k})^{-1/2},
\nonumber\\
\tilde{q}_{{\bf k}}&&\ =\ (-\cos(k_{y}/2)p_{{\bf k}x}
+\cos(k_{x}/2) p_{{\bf k}y})(1+\gamma_{\bf k})^{-1/2},
\nonumber
\end{eqnarray}
where $\gamma_{{\bf k}}$=(1/2)(cos(k$_{x}$a)+
cos(k$_{y}$a)).

Thus, the $pd$- hybridization term in $H'$ includes only the
symmetrical state \cite{Zh1}. In terms of these  operators (\ref{5}) we
rewrite the Hamiltonian (\ref{3}), (\ref{4}) as:
\begin{eqnarray}
\label{7}
H_{0}&&=\epsilon_{d}\sum_{{\bf k},\alpha}
d^{+}_{{\bf k}\alpha}d_{{\bf k}\alpha}+
\epsilon_{p}\sum_{{\bf k},\alpha}
(q^{+}_{{\bf k}\alpha}q_{{\bf k}\alpha}
+\tilde{q}^{+}_{{\bf k}\alpha}\tilde{q}_{{\bf k}\alpha})
\nonumber\\
&&+U_{d}\sum_{\bf k_{1}+k_{2}=k_{3}+k_{4}}
d^{+}_{{\bf k_{1}}\uparrow}d_{{\bf k_{2}}\uparrow}
d^{+}_{{\bf k_{3}}\downarrow}d_{{\bf k_{4}}\downarrow} \ ,
\\
H'&&=2t\sum_{{\bf k}\alpha} \lambda _{{\bf k}}
\{ d^{+}_{{\bf k}\alpha} q_{{\bf k}\alpha} + H.c.\}
\nonumber\\
&&-t_{p}\sum_{{\bf k}\alpha} \{ \mu _{{\bf k}} [
q^{+}_{{\bf k}\alpha} q_{{\bf k}\alpha}-
\tilde{q}^{+}_{{\bf k}\alpha} \tilde{q}_{{\bf k}\alpha}]+
\nu _{{\bf k}} [q^{+}_{{\bf k}\alpha} \tilde{q}_{{\bf k}\alpha}+
\tilde{q}^{+}_{{\bf k}\alpha} q_{{\bf k}\alpha} ] \}
\nonumber
\end{eqnarray}
were quantities $\lambda_{\bf k},\ \mu_{\bf k},\ \nu_{\bf k}$  are
\begin{eqnarray*}
\lambda_{\bf k}&&=(1+\gamma_{\bf k})^{1/2} \ ,\ \
\nonumber\\
\mu_{\bf k}&&=8\cos^{2}(k_{x}/2) \cos^{2}(k_{y}/2)
(1+\gamma_{\bf k})^{-1/2} \ , \\
\nu_{\bf k}&&=4\cos(k_{x}/2) \cos(k_{y}/2)(\cos^{2}(k_{x}/2)
\nonumber\\
&&- \cos^{2}(k_{y}/2))(1+\gamma_{\bf k})^{-1/2} \ .
\nonumber
\end{eqnarray*}
It is not convenient to treat   this Hamiltonian in the ${\bf k}$-space
due to the strong hole repulsion at the copper sites or large value
of the parameter U$_{d}$ in this Hamiltonian. For taking into
account this term precisely, we return to lattice representation for
our Hamiltonian.  After inverse Fourier transformation, Hamiltonian
(\ref{7})  has the form:
\begin{eqnarray}
\label{8}
H_{0}&&=\epsilon_{d}\sum_{l,\alpha} d^{+}_{l\alpha}d_{l\alpha}
+\epsilon_{p}\sum_{l,\alpha}(q^{+}_{l\alpha}q_{l\alpha}+
\tilde{q}^{+}_{l\alpha} \tilde{q}_{l\alpha})
\nonumber\\
&&+U_{d}\sum_{l}d^{+}_{l\uparrow}d_{l\uparrow}d^{+}_{l\downarrow}
d_{l\downarrow}
\nonumber\\
H'&&=2t\sum_{\langle ll'\rangle\alpha}\lambda_{ll'}
(d^{+}_{l\alpha}q_{l'\alpha}+ H.c.)
\\
&&-t_{p}\sum_{\langle ll'\rangle\alpha}
\{\mu_{ll'}(q^{+}_{l\alpha}q_{l'\alpha}-
\tilde{q}^{+}_{l\alpha} \tilde{q}_{l'\alpha})
\nonumber\\
&&+\nu_{ll'}(q^{+}_{l\alpha}\tilde{q}_{l'\alpha}+ H.c)\} \ ,
\nonumber
\end{eqnarray}
where
\begin{eqnarray}
\label{9}
&&(q_{l},\tilde{q}_{l})=
 \sum_{\bf k} (q_{{\bf k}},\tilde{q}_{{\bf k}})
 \exp(-i{\bf k}({\bf r_{l}})) \ ,
\nonumber\\
&&\{\lambda, \mu, \nu\}_{ll'} \equiv \{\lambda, \mu, \nu\}({\bf l}-{\bf l'})
\\
 &&=\sum_{\bf k}\{\lambda, \mu, \nu\}_{\bf k} \exp(-i{\bf k}({\bf l-l'})) \ ,
\nonumber
\end{eqnarray}
where summation over ${\bf k}$ is produced over the Brillouin zone,
coefficients $\lambda, \mu, \nu$ decrease rapidly with increasing
$|{\bf l}-{\bf l'}|$. The
values of $\lambda$, $\mu$ and $\nu$ for small $|{\bf l}-{\bf l'}|$ are
given in Table I.
It is easy to get that $\nu_{0} \equiv\nu _{00} =0 $ and $\nu
_{n,n}\equiv \nu (n{\bf x} +n{\bf y}) =0 $.

One can divide Hamiltonian (\ref{8}) into the local and hopping parts
\begin{eqnarray}
\label{10}
H_{loc}&&=\epsilon_{d}\sum_{l,\alpha} d^{+}_{l\alpha}d_{l\alpha}+
(\epsilon_{p}-\mu_{0}t_{p})\sum_{l,\alpha}q^{+}_{l\alpha}q_{l\alpha}
\nonumber\\
&&+(\epsilon_{p}+\mu_{0}t_{p})\sum_{l,\alpha}\tilde{q}^{+}_{l\alpha}
 \tilde{q}_{l\alpha}
\nonumber\\
&&+U_{d}\sum_{l}d^{+}_{l\uparrow}d_{l\uparrow}
d^{+}_{l\downarrow}d_{l\downarrow}
\\
&&+2t\lambda_{0}\sum_{ll'\alpha}(d^{+}_{l\alpha}q_{l\alpha}+ H.c.) \ ,
\nonumber\\
H_{hop}&&=2t\sum_{ll'\alpha}\lambda_{ll'}(d^{+}_{l\alpha}q_{l'\alpha}+ H.c.)
\nonumber\\
&&-t_{p}\sum_{ll'\alpha}\{\mu_{ll'}(q^{+}_{l\alpha}q_{l'\alpha}-
\tilde{q}^{+}_{l\alpha} \tilde{q}_{l'\alpha})
\nonumber\\
&&+\nu_{ll'}(q^{+}_{l\alpha}\tilde{q}_{l'\alpha}+ H.c.)\} \ ,
\nonumber
\end{eqnarray}
hereafter sum over $l,l'$ means that $l\neq l'$.
Now, one can discuss the reasons for transformation from
the Hamiltonian (\ref{1}) - (\ref{2}) to (\ref{10}) . The copper
$d$-state hybridizes only with the symmetrical oxygen $q$-state
\cite{Zh1}. Taking into account the direct O-O hopping does not
change this picture.  After this separation out of the strongly
interacting $d$ and $q$ local states, one can find the full energy
spectrum of the one- or two-hole local states and solve the problem
of the low-energy two-hole state at the background of the one-hole
states (spins) consistently.  The direct O-O hopping only slightly
shifts the energies of the oxygen states with the opposite symmetry.
But its contribution to the effective hopping parameters will be very
important (see Sec. V).  The transformed Hamiltonian (\ref{10}) with the
definitions (\ref{9}) is equivalent to the Hamiltonian
(\ref{1}), (\ref{2}) without the additional Coulomb terms U$_{p}$
and V$_{pd}$ and describes the three-band model exactly.

\section{ Diagonalization of the local Hamiltonian} \ \ \
In this paper we study the model  relating to high-temperature
superconductors, which are the systems with a nearly half-filled band.
Thus, we shall concentrate our attention on the case of one hole over
unit filling in the framework of the three-band model.  Namely, we
determine the space of one- and two-hole local states of the CuO$_{2}$
plane.  A primary set of states in the one-hole sector is
\begin{eqnarray}
\label{11}
&&|d_{\alpha}\rangle\equiv d^{+}_{\alpha}|0\rangle \  ,\
|q_{\alpha}\rangle\equiv q^{+}_{\alpha}|0\rangle \  ,  \
\nonumber\\
&&|\tilde{q}_{\alpha}\rangle\equiv \tilde{q}^{+}_{\alpha}|0\rangle \ .
\end{eqnarray}
The two-hole singlet sector has a set of states
\begin{eqnarray}
\label{12}
&&|\psi\rangle\equiv d^{+}_{\uparrow }d^{+}_{\downarrow }|0\rangle \  ,\
|\varphi \rangle\equiv q^{+}_{\uparrow }q^{+}_{\downarrow }|0\rangle \  ,\
|\chi \rangle\equiv (1/\sqrt{2})(d^{+}_{\uparrow }q^{+}_{\downarrow }-
d^{+}_{\downarrow }q^{+}_{\uparrow })|0\rangle \  ,\
\nonumber \\
&&|\tilde{\varphi}\rangle\equiv \tilde{q}^{+}_{\uparrow }
\tilde{q}^{+}_{\downarrow }|0\rangle \  ,\
|\tilde{\chi} \rangle\equiv (1/\sqrt{2})(d^{+}_{\uparrow }
\tilde{q}^{+}_{\downarrow }-
d^{+}_{\downarrow }\tilde{q}^{+}_{\uparrow })|0\rangle \  ,\  \\
&&|\tilde{\eta } \rangle\equiv (1/\sqrt{2})(q^{+}_{\uparrow }
\tilde{q}^{+}_{\downarrow }-
q^{+}_{\downarrow }\tilde{q}^{+}_{\uparrow })|0\rangle \  ,\ \nonumber
\end{eqnarray}
The two-hole triplet sector is
\begin{eqnarray}
\label{13}
&&|\tau \rangle\equiv\left(d^{+}_{\uparrow } q^{+}_{\uparrow },
 d^{+}_{\downarrow }q^{+}_{\downarrow },
 (1/\sqrt{2})(d^{+}_{\uparrow }q^{+}_{\downarrow }+
d^{+}_{\downarrow }q^{+}_{\uparrow })\right)|0\rangle \  ,
\nonumber\\
&&|\tilde{\tau} \rangle\equiv \left(d^{+}_{\uparrow }
\tilde{q}^{+}_{\uparrow },
 d^{+}_{\downarrow }\tilde{q}^{+}_{\downarrow },
 (1/\sqrt{2})(d^{+}_{\uparrow }\tilde{q}^{+}_{\downarrow }+
d^{+}_{\downarrow }\tilde{q}^{+}_{\uparrow })\right)|0\rangle \  ,
\\
&&|\zeta \rangle\equiv \left(q^{+}_{\uparrow } \tilde{q}^{+}_{\uparrow },
 q^{+}_{\downarrow }\tilde{q}^{+}_{\downarrow },
 (1/\sqrt{2})(q^{+}_{\uparrow }\tilde{q}^{+}_{\downarrow }+
q^{+}_{\downarrow }\tilde{q}^{+}_{\uparrow })\right)|0\rangle  \  ,
\nonumber
\end{eqnarray}
where $|0\rangle$ is the vacuum state (state without holes on a
cluster).  Hence, we have three one-hole and nine two-hole states.
As will be shown below, some of these states do not play any role
in a low-energy model.  In this part of the work we shall consider only
the local part $H_{loc}$ of the Hamiltonian (\ref{10}).  It is
convenient to express this Hamiltonian in terms of the Hubbard
operators rather than in terms of the usual Fermi operators.  Such
representation of the Hamiltonian (\ref{10}) in the one- and two-hole
sector in terms of states (\ref{11}-\ref{13}) has the form:
\begin{eqnarray}
\label{14}
H^{1}_{loc}&&=\sum_{l,\alpha} \{ \epsilon_{d}
X^{d\alpha d\alpha }_{l}+(\epsilon_{p}-\mu_{0}t_{p})
X^{q\alpha q\alpha }_{l}
\nonumber\\
&&+(\epsilon_{p}+\mu_{0}t_{p}) X^{\tilde{q}\alpha\tilde{q}\alpha }_{l}
\nonumber\\
&&+2t\lambda_{0}(X^{d\alpha q\alpha}_{l} + H.c.) \} \ ,
\nonumber\\
H^{2}_{loc}&&=\sum_{l} \left\{ \left[ (U_{d}+2\epsilon_{d})
X^{\psi \psi}_{l}+2(\epsilon_{p}-\mu_{0}t_{p}) X^{\varphi \varphi }_{l}
\right.\right.
\nonumber\\
&&\left.
+ (\epsilon_{d}+\epsilon_{p}-\mu_{0}t_{p}) X^{\chi \chi}_{l}
+2\sqrt{2}t\lambda_{0}(X^{\chi \psi}_{l} +X^{\chi\varphi}_{l}
 + H.c.)\right]
\\
&&+2(\epsilon_{p}+\mu_{0}t_{p}) X^{\tilde{\varphi}\tilde{\varphi}}_{l}
 +\left[(\epsilon_{d}+\epsilon_{p}+\mu_{0}t_{p})
X^{\tilde{\chi}\tilde{\chi}}_{l}+\epsilon_{p}X^{\eta \eta}_{l}
\right.
\nonumber\\
&& \left.
+2t\lambda_{0}(X^{\tilde{\chi}\eta}_{l} + H.c.)\right]
+ (\epsilon_{d}+\epsilon_{p}-\mu_{0}t_{p}) \sum_{m =\pm1,0}
X_{l}^{\tau m \tau m}
\nonumber\\
&&+\left. \sum_{m =\pm1,0} \left[(\epsilon_{d}+\mu_{0}t_{p})
X_{l}^{\tilde{\tau} m\tilde{ \tau} m}
+\epsilon_{p} X_{l}^{\zeta m\zeta m}
+2t\lambda_{0} (X_{l}^{\zeta m \tilde{\tau } m}
 + H.c.)\right] \right\}\ ,
\nonumber
\end{eqnarray}
where the upper index 1,2 marks one- and two-hole sector of the local
Hamiltonian, $X_{l}^{ab}$ is the Hubbard operator at the site $l$:
\begin{eqnarray}
\label{14a}
X^{a b}_{l} \equiv |a\ l\rangle\langle b \ l| \ , \ \
a,\ b \ = \ \{ d\alpha ,\ q\alpha ,\ \tilde{q}\alpha ,\ \psi ,\
\varphi ,\ \chi ,\ \tilde{\chi },\ \eta ,\ \tau m,\
\zeta m,\ \tilde{\tau }m \},
\end{eqnarray}
where index $\alpha =\pm1/2$ in $H^{1}$ is a spin projection, $m\equiv\pm1,0$
in $H^{2}$ denotes triplet components.

The representation of $H_{loc}$ (\ref{10}) in terms of the Hubbard
operators Eq. (\ref{14}) allows us to solve simply the one-site problem.
The diagonalization of $H_{loc}^{1}, H_{loc}^{2}$ is performed at
each site independently.
After the diagonalization, $H_{loc}^{1}$ is given by
\begin{eqnarray}
\label{15}
H^{1}_{loc}&&=\sum_{l,\alpha} \left\{ \epsilon_{f}
X^{f\alpha f\alpha }_{l}+\epsilon_{g} X^{g\alpha g\alpha }_{l} \right.
\nonumber\\
&&\left. +(\epsilon_{p}+\mu_{0}t_{p})
X^{\tilde{q}\alpha  \tilde{q}\alpha }_{l} \right\} \ ,
\end{eqnarray}
where
\begin{eqnarray}
\label{15a}
&&\epsilon_{f,g}=-\Delta_{1} \mp R_{1}\ ,\
\Delta_{1}=(\Delta+\mu _{0}t_{p})/2 \ ,
\nonumber\\
&&R_{1}=(\tilde{\Delta}^{2} + 4t^{2}\lambda_{0}^{2})^{1/2} ,\
\tilde{\Delta}=(\Delta-\mu_{0}t_{p})/2 ,\
\Delta=\epsilon_{p}-
\epsilon_{d}\ ,
\end{eqnarray}
$|f\alpha \rangle $ and $|g\alpha \rangle $ are the lower and
higher one-hole cluster states
\begin{eqnarray}
\label{16}
&&|f\alpha\rangle=U |d\alpha\rangle - V |q\alpha\rangle \ ,
\nonumber\\
&&|g\alpha\rangle=V |d\alpha\rangle + U |q\alpha\rangle \ ,
\\
&&U=((R_{1}+\tilde{\Delta})/2R_{1})^{1/2} \ ,
\nonumber\\
&&V=((R_{1}-\tilde{\Delta})/2R_{1})^{1/2} \ ,
\nonumber
\end{eqnarray}
Our approach gives us the reasons to assume that in the low-energy
limit at unit filling the background of the CuO$_{2}$ plane consists
of the lowest $|f\alpha l\rangle$-states (\ref{16}) at each cluster.
Such state is a linear combination of the $d$-hole state and
the orthonormalized symmetrical $q$-hole state at the nearest oxygens.
Virtual transitions of holes in the $|f\alpha \rangle$-states (spins) at the
neighbor sites and back provide the antiferromagnetic type of
interaction between these spins.

In $H_{loc}^{2}$ (\ref{14}) there are different sectors, separated by
the square brackets, which are diagonalized independently.  The first
sector includes the three singlets $dd$, $qq$, and $dq$.  One can prove
that diagonalization of this sector gives the lowest singlet with
the eigenenergy well below than the eigenenergies of other
states.  Thus, the diagonalized two-hole part of the Hamiltonian
(\ref{14}) is
\begin{eqnarray}
\label{17}
H^{2}_{loc}=\sum_{l} \left\{ E_{-}
X^{cc}_{l}+\sum_{y} E_{-} X^{yy}_{l} \right\} ,
\end{eqnarray}
where $E_{-}$ is the energy of the lowest singlet $|c\rangle$, $E_{y}$ are
the energies of the upper states, $y$ is the set of the two-hole
local states without $|c\rangle$.  For the analytical expression of the
eigenenergies and for the set of eigenstates see Sections V, VI.
After diagonalization of the local part (\ref{14}) of the full
Hamiltonian (\ref{10}), one can introduce the nondiagonal Hubbard
operators and rewrite the hopping Hamiltonian (\ref{10})  in terms of
transitions between local eigenstates at different sites.

\section{Low-energy reduction}
\subsection{ Zero order} \ \
As was noted, we assume that at unit filling there are holes in the
lowest $|f\alpha \rangle$-states (spins) at each cluster.   We will
demonstrate below that these spins interact antiferromagnetically.
A general expression for the superexchange constant $J$ is derived in
Section IV.D.

For the case of one hole over unit filling we take into account only the
lowest singlet and its transitions from site to site.  All upper
two-hole states and transitions to them are projected out.  Such
procedure gives the $t$-$t'$-$J$ model with hopping at all
neighbors in the explicit form:
\begin{eqnarray}
\label{18}
H_{t-t'}= E_{-}\sum_{l} X^{cc}_{l}+
\sum_{ll'\alpha}t_{ll'}X^{f\alpha,c}_{l'}X^{c,f\alpha }_{l} \ ,
\end{eqnarray}
where $|c\rangle$ is the lowest singlet, $|f\alpha \rangle$ is the
lowest one-hole state, $t_{ll'}$ are the effective hopping
parameters from $l$ to $l'$ site.  In more usual terms this
Hamiltonian (\ref{18}) may be written as
\begin{eqnarray}
\label{19}
&&H_{t-t'}=E_{-}\sum_{l}(1-\hat{n}_{l}^{c}+\hat{d}_{l}^{c})+
\sum_{ll',\alpha}t_{ll'} \tilde{c}^{+}_{l\alpha} \tilde{c}_{l'\alpha}
\nonumber\\
&&\tilde{c}_{l\alpha} =c_{l\alpha}(1-\hat{n}_{l,-\alpha}) \ , \ \
\tilde{c}^{+}_{l\alpha}=(\tilde{c}_{l\alpha})^{+} \ , \ \
\hat{n}_{l}=(c_{l}^{+}\cdot c_{l}) \ ,\ \
\hat{d}_{l}^{c}\equiv
 c^{+}_{\uparrow}c_{\uparrow}c^{+}_{\downarrow}c_{\downarrow}.
\end{eqnarray}
Here $c_{l\alpha}^{+},c_{l\alpha}$  are the electron creation
 and annihilation operators.

In the next part of the work we will consider the second-order perturbation
theory, which gives us a criterion of validity of the $t$-$t'$-$J$
model (\ref{18}).  Thus, if corrections to the energy and the hopping
integrals for the lowest singlet from virtual transitions to the upper
states are small, one can argue that this model (\ref{18}) is
valid.

\subsection{ Schrieffer-Wolff transformation}\ \
By diagonalization of the local part of the primary Hamiltonian
(\ref{14})  we obtain a set of the local states.  For the low-energy
processes one can consider only the lowest states and include all
others perturbatively.   We use the Schrieffer-Wollf (S-W)
transformation \cite{Bi1}
\begin{eqnarray}
\label{20}
H \Rightarrow \tilde{H}=\exp{(-S)} H \exp{(S)}\ , \
 S^{+}=-S .
\end{eqnarray}
for getting the second-order correction to the pure $t$-$t'$-$J$
model.  On this way we use the smallness of the $\{\lambda  ,\mu , \nu
\}_{ll'}$ constants for  $|{\bf l} -{\bf l'}| \neq 0$.
The first-order generator of transformation  $S$  and second-order
correction are given by
\begin{eqnarray}
\label{21}
[H_{0},S]=-H', \ \ \ \delta H^{2}=(1/2)[H',S] \ ,
\end{eqnarray}
For the case of one hole over unit filling $H_{0} \equiv H_{loc}^{2}
+H_{loc}^{1}+H_{t-t'}$ (\ref{15},\ref{17},\ref{18}) and $H'$ includes all
terms of $H_{hop}$ (\ref{10}) which are relevant to the transition
between the lowest singlet and all of the upper two-holes states.
In terms of the nondiagonal Hubbard operators the Hamiltonian
$H'$ is given by
\begin{eqnarray}
\label{22}
H_{hop}&&=\sum_{ll'\alpha \beta}\sum_{y}
\left.' \right.F^{f\beta ,y}_{ll',\alpha }\left\{
X^{y,f\alpha}_{l}X^{f\beta,c}_{l'} + H.c. \right\}
\nonumber\\
&&+\sum_{ll'\alpha\beta ,y} F^{g\beta,y}_{ll',\alpha}\left\{
X^{y,f\alpha}_{l}X^{g\beta,c}_{l'}+ H.c.\right\} \ ,
\end{eqnarray}
where $y$ is the set of the two-hole states, the prime in the sum (\ref{22})
denotes absence of contribution of the lowest singlet state;
$f\alpha $ and $g\beta $ are the one-holes states (\ref{16}).
The second term in Eq.(\ref{22}) is essential only for correction to
the energy $E_{-}$ of the $c$-singlet.  The first term in
Eq.(\ref{22}) provides corrections both to the energy and to the hopping
integrals.  $F_{ll'\alpha} ^{x\beta ,y} $ are  the matrix
elements of transitions between $|cl'\rangle \otimes |f\alpha l\rangle$
and $|x\beta l'\rangle  \otimes |yl\rangle$ states ($x=f,g$).  Their
explicit form is presented in Appendix A.

In the case of unit filling, $H_{0}$ for Eq.(\ref{21}) is
\begin{eqnarray*}
H_{0} \equiv H_{loc}^{2}+H_{loc}^{1}
\end{eqnarray*}
and $H'$ is
\begin{eqnarray}
\label{24}
H'&&=\sum_{ll'\alpha\beta ,y}D^{y}_{ll',\alpha \beta }\left\{
X^{y,f\alpha}_{l}X^{0,f\beta }_{l'}+ H.c.
\right\} \ ,
\end{eqnarray}
where $|0\rangle$ is the state without holes, $D_{ll',\alpha \beta}
^{y}$ are the matrix elements of transition of two holes at
different sites into the two-holes state $y$ :
$|f\alpha l\rangle  \otimes |f\beta l'\rangle \Rightarrow
|0l\rangle  \otimes |yl'\rangle$ (see Appendix B).
Corresponding generators of the S-W transformation are given in
Appendices A,B.

\subsection{Second-order corrections}\ \
Here we derive a general form of corrections to the $t$-$t'$-$J$
model.  By applying the S-W transformation (\ref{20}), (\ref{21}) to
the Hamiltonian $H'$ (\ref{22}) one can get correction to the
Hamiltonian of $c$-singlet \cite{Be1}
\begin{eqnarray}
\label{25}
\delta H_{E}=-\sum_{ll'}M_{ll'}X^{cc}_{l}\hat{N}_{l'}
\end{eqnarray}
with
\begin{eqnarray}
\label{26}
M_{ll'}=\sum_{y}\left.'\right. \frac{|F_{ll'}^{f,y}|^{2}}{E_{y}-E_{-}}+
\sum_{y} \frac{|F_{ll'}^{g,y}|^{2}}{E_{y}-E_{-}+\epsilon_{g}- \epsilon_{f}}
, \
\end{eqnarray}
and $\hat{N}_{l} =X_{l}^{\uparrow \uparrow }+
 X_{l}^{\downarrow \downarrow }$.

Corrections to the hopping Hamiltonian have the form \cite{Be1}
\begin{eqnarray}
\label{27}
\delta H_{t}&&=\sum _{lnl'\alpha \beta}\sum_{y}\left.'\right.
T_{lnl'}^{y}\left\{x_{y}
X^{f\alpha, c}_{l}X^{c,f\alpha}_{l'}\hat{N}_{n} \right.
\nonumber \\
&&\left.
+z_{y}X^{f\alpha ,c}_{l}X^{c,f\beta}_{l'}
({\mbox{\boldmath $\sigma$}}_{ \alpha \beta}{\bf S}_{n})\right\} \ ,
\\
\mbox{where}
\nonumber \\
&&{\bf S}_{l}=(1/2){\mbox{\boldmath $\sigma$}}_{\alpha \beta }
X_{l}^{\alpha \beta }
\ ,
\nonumber
\end{eqnarray}
here $x_{y}=1/2 $ for singlets and $x_{y}=3/4$ for triplets; $z_{y}=1$
for singlets and $z_{y}=-1/2$ for triplets
\begin{eqnarray}
\label{28}
T_{lnl'}^{y}=\frac{F_{ln}^{y}F_{nl'}^{y}}{E_{y}-E_{-}} \ .
\end{eqnarray}
In Eqs. (\ref{26}), (\ref{28}) we express parameters $M$ and $T$
through the matrix elements without spin structure.  They are connected
with the primary matrix element as
\begin{eqnarray}
\label{29}
F_{ll',\alpha}^{x\beta ,y}=\delta_{\alpha \beta } F_{ll'}^{x,y} \ ,
\end{eqnarray}
for $y$=singlet and
\begin{eqnarray}
\label{30}
F_{ll',\alpha}^{x\beta ,y}=(2\beta ) [m\alpha \bar{\beta }] F_{ll'}^{x,y} \ ,
\end{eqnarray}
for $y$= triplet, where $[m\alpha \bar{\beta}] = \langle 1/2\alpha
| 1/2 \bar{\beta } ,1 m\rangle$ are the
Clebsh-Cordon coefficients , $m=\pm1,0$ are spin-$1$
projections, $\alpha ,\beta =\pm 1/2$ are the spin-$\frac{1}{2}$ projections,
$\bar{\beta }\ =\ -\beta $.
One can see that the corrections (\ref{25}), (\ref{27}) to the
$t$-$t'$-$J$ Hamiltonian (\ref{18}) depend on the filling
$(\hat{N}$-term in $\delta H_{E} $  and $\delta H_{t})$ and magnetic
order (${\bf S}$ term in $\delta H_{t}$ ).  Such terms can not be
expressed through the simple direct hopping.

In Section V  we will analyze quantitatively the relative magnitudes
of the second-order corrections.  Validity of the $t$-$J$ model, as well as
correctness of inclusion of hopping at the next neighbors are
well checked by this analysis.

\subsection{ Superexchange  interaction}\ \
In our approach the superexchange interaction arises in the second
order of the perturbation theory over hopping of holes at neighboring
cluster and back.  The situation is similar to calculation of the
superexchange interaction in the simple Hubbard
model ($4t^{2}/U$).  Of course, in
the case of small $t$  our result for the superexchange constant must
be proportional to $t^{4}$ as was calculated in earlier works
\cite{Zh2,Em2} in the fourth order of perturbation theory.  From Eq.
 (\ref{24})
one can get
\begin{eqnarray}
\label{31}
\delta H^{2}=H_{J}=\sum_{ll'}(J_{ll'}{\bf S}_{l}{\bf S}_{l'}
+Y_{ll'}\hat{N}_{l}\hat{N}_{l'}) \ .
\end{eqnarray}
Such form (\ref{31}) is more general than in the $t$-$J$ model.  Firstly,
there are interactions between all pairs of spins.
Due to the rapid decrease of the constant $\{\lambda ,\mu ,\nu \}$,
 ($J_{\langle ll'\rangle_{2,3}} \sim 10^{-2} J_{\langle ll'\rangle}$), one
 can omit all
next-nearest neighbor terms in Eq. (\ref{31}) and get the Heisenberg term
of the usual $t$-$J$ Hamiltonian. Secondly, $Y_{ll'}\neq
-(1/4)J_{ll'}$ because the hole may virtually hop into triplet states
and back, i.e.
\begin{eqnarray}
\label{33}
J_{ll'}&&=\sum_{y}x_{y}\frac{|D_{ll'}^{y}|^{2}}{\epsilon_{y}+\epsilon _{0}
- 2\epsilon_{f}} \ , \
\nonumber\\
Y_{ll'}&&=-\sum_{y}z_{y}\frac{|D_{ll'}^{y}|^{2}}{\epsilon_{y}+\epsilon _{0}
- 2\epsilon_{f}} \ , \
\end{eqnarray}
with $x_{y}=2 $ and $z_{y}=1/2$  for $y$ =singlet, $x_{y} =-1$ and
$z_y=+3/4$ for $y$=triplet.
$D_{ll'}^{y}$ are connected with the matrix elements
$D_{ll'\alpha \beta } ^{y} $ as follows
\begin{eqnarray}
\label{34}
D_{ll',\alpha \beta }^{y}=(2\alpha )\delta_{\alpha \bar{\beta}}
D_{ll'}^{y} \
\end{eqnarray}
for $y$=singlet,
\begin{eqnarray}
\label{35}
D_{ll',\alpha \beta }^{y}=[m\alpha \beta] D_{ll'}^{y} \
\end{eqnarray}
for $y$ =triplet.
One can check that in the limit $t\Rightarrow 0$ the reduced matrix
element $D_{ll'}^{y}$ is proportional to $t$ and the superexchange
constant $J_{ll'}$ has contribution proportional to $t^{2}$.  But
such contributions from the singlet and the triplet two-hole states
are cancelled, and we get usual result which is proportional to
$t^{4}$.

\section{Limiting cases}\ \ \
In the previous parts of the work we consistently reformulated the
three-band model and passed from the terms of the primary Fermi operators of
the hole at the copper and at the oxygens to the one- and two-holes states
Hubbard operators on the cluster.
For the problem of one hole over unit filling we established the
general form of the low-energy the  $t$-$t'$-$J$ model
(\ref{18}), the general form of the corrections to it
(\ref{25}), (\ref{27}), and  the general form of the spin interaction at
unit filling (\ref{31}). Now we will consider three limiting cases,
all with special constraints on the Gilbert space of the problem.
Such constraints simplify the mathematical treatments and provide the
possibility to get results in the analytical form.

\subsection{ The case of $\Delta \ll U_{d}$ }\ \
This case was considered by Lovtsov and Yushankai \cite{Yu1}. We
include into this problem direct O-O hopping and make
consideration more complete.

Such limit (U$_{d} \Rightarrow \infty $) pushes up in energy the
state with double occupation of the copper.  Thus, one can ignore
$|\psi\rangle$-singlet state (\ref{12}).  After such simplification
diagonalization of the two-particle sector of the Hamiltonian
(\ref{14}) presents no problems
\begin{eqnarray}
\label{36}
H^{2}_{loc}&&=\sum_{l} \left\{ E_{-}X^{cc}_{l}+E_{+} X^{bb}_{l}
\right.
\nonumber\\
&&+2\mu_{0}t_{p} X^{\tilde{\varphi}\tilde{\varphi}}_{l}
 +\tilde{E}_{-}\left( X^{\tilde{c}\tilde{c}}_{l}
+ \sum_{m =\pm1,0}X_{l}^{t1m\ t1m}\right)
\\
&&\left.
 +\tilde{E}_{+}\left( X^{\tilde{b}\tilde{b}}_{l}
+ \sum_{m =\pm1,0}X_{l}^{t2m\ t2m}\right)-
 (\Delta +\mu_{0}t_{p}) \sum_{m =\pm1,0} X_{l}^{\tau m\ \tau m}
 \right\}\ ,
\nonumber
\end{eqnarray}
where $c,b$ are the lowest and highest singlets of $q-d$ sector,
$\tilde{c},\tilde{b}$ are the lowest and highest $\tilde{q}-d$ singlets,
$t1,t2$ are the lowest and highest $\tilde{q}-d$ triplets, $\tau $
and $\tilde{\varphi} $ are determined
in Eqs. (\ref{12}), (\ref{13}).  All energies are counted from $\epsilon_{p}$
 (hereafter $\epsilon _{p}=0$).
$E_{\pm} = -\Delta_{2}\pm R$,  $\Delta _{2}=1/2(\Delta  +3\mu
_{0}t_{p})$,  $R =(\tilde{\Delta }^{2} +8t^{2}\lambda _{0}^{2})^{1/2}$,
 $\tilde{E}_{\pm} =-\tilde{\Delta} \pm R_{1}$,
 $\tilde{\Delta}=(\Delta-\mu_{0}t_{p})/2$.  New eigenstates are connected
with the primary set of states (\ref{12}), (\ref{13}), as
\begin{eqnarray}
\label{37}
&&|c\rangle= - a |\varphi \rangle + b |\chi \rangle \ ,\
|b\rangle= b |\varphi \rangle + a |\chi \rangle \ ,
\nonumber\\
&&b=((R+\tilde{\Delta})/2R)^{1/2} \ ,\
a=((R-\tilde{\Delta})/2R)^{1/2} \ ,
\\
&&|\tilde{c}\rangle= U|\tilde{\chi}\rangle - V |\tilde{\eta } \rangle\ ,\
|\tilde{b}\rangle= V|\tilde{\chi}\rangle + U |\tilde{\eta } \rangle\ ,\
\nonumber\\
&&|t1\rangle = U|\tilde{\tau }\rangle - V |\zeta \rangle \ ,\
|t2\rangle= V|\tilde{\tau }\rangle +U |\zeta \rangle \ ,\
\end{eqnarray}
$U$ and $V$ are determined in Eq.(\ref{16})  The energies of the local
states at U$_{d} \Rightarrow \infty$ (\ref{36}) for values of the
hopping parameters $t=1.4$ eV, $t_{p} =0.7$ eV and $\Delta =3.66$ eV are
given in Table II.

Next, we follow the scheme of Sec. III.  Explicit forms of
parameters of the $t$-$t'$-$J$ model are
\begin{eqnarray}
\label{38}
E_{c}&&=E_{-}=-(\Delta+3\mu_{0}t_{p})/2 -R
\nonumber\\
t_{ll'}&&=2t\lambda _{ll'}V b(bU+\sqrt{2}V a)
+ t_{p}\mu _{ll'}
(\sqrt{2} a V + U b)^{2}/2  \ ,
\end{eqnarray}
Hopping parameters for the first four neighbors at values of
parameters mentioned above are given in Table II. One can see the
important role of the direct O-O hopping.  Its contribution to the
full amplitude of hopping at the nearest neighbor $\approx 50\%$.
Explicit  form of the matrix elements $F_{ll'}^{yx}$ and
$D_{ll'}^{y}$ one can get from Eqs.
(\ref{37}), (\ref{22}), (\ref{24}), and (\ref{10}).  They are given in
Appendices A and B.  Relative magnitudes  of corrections to the
hopping and energy are presented in Tab. II.  Corrections to the energy
and to the first hopping parameter are $4.1\%$  and $3.1\%$,  respectively.
Thus, the $t$-$J$ model is valid for this case.  It seems that inclusion of
hopping at the second and third neighbor to the $t$-$t'$-$J$ model is
justified because  corrections to them are small enough.

\subsection{The case (U$_{d}-\Delta) \ll $U$_{d}$ }\ \
This case was considered by the authors \cite{Be1} without the O-O direct
hopping.  In Ref. \cite{Be1} we assumed that at unit filling there is
one hole (spin) per copper site.  In terms of the present work we
ignored the admixture of the $|q\rangle$-state to $|d\rangle$-state.
We also assumed that there are no doubly occupied oxygen degrees of
freedom.  Both assumptions are valid over the parameter $\sim t/U_{d}$.
Such constraints lead to $U$ =1,  $V$=0 (see Eq.(\ref{16})) and rather
simple form of the diagonalized Hamiltonian $H_{loc}^{2}$
\begin{eqnarray}
\label{39}
H^{2}_{loc}&&=\sum_{l} \left\{ E_{-}X^{cc}_{l}+E_{+} X^{bb}_{l}
- (\delta +\mu_{0}t_{p}) \sum_{m =\pm1,0} X_{l}^{\tau m \ \tau m}
\right.
\nonumber\\
&&\left.
 -(\delta -\mu_{0}t_{p})\left( X^{\tilde{\chi }\tilde{\chi }}_{l}+
 \sum_{m =\pm1,0} X_{l}^{\tilde{\tau} m \ \tilde{\tau} m}\right)
 \right\}\ ,
\nonumber
\end{eqnarray}
where
$E_{\pm} = -\Delta_{1}\pm R$,  $\Delta _{1}=1/2(U_{d}-3\Delta -\mu
_{0}t_{p})$,  $R =(\tilde{\Delta }^{2} +8t^{2}\lambda _{0}^{2})^{1/2}$,
 $\tilde{\Delta}=(U_{d}-\Delta+\mu_{0}t_{p})/2$,
     and
\begin{eqnarray}
\label{40}
&&|c\rangle= - a |\psi \rangle + b |\chi \rangle \ ,\
|b\rangle= b |\psi \rangle + a |\chi \rangle \ ,
\nonumber\\
&&b=((R+\tilde{\Delta})/2R)^{1/2} \ ,\
a=((R-\tilde{\Delta})/2R)^{1/2} \ .
\end{eqnarray}
The explicit form of parameters of the $t$-$t'$-$J$ model is following
\begin{eqnarray}
\label{41}
E_{c}&&=E_{-}=-(U_{d}-3\Delta-\mu_{0}t_{p})/2 -R
\nonumber\\
t_{ll'}&&=2t\lambda _{ll'}\sqrt{2} b a
+ t_{p}\mu _{ll'} b^{2}/2  \ .
\end{eqnarray}
The spectrum of the local energies and first four hopping integrals
at U$_{d} =8$ eV and $\Delta  =3.65$ eV  are given in Table III.
One can see that direct O-O hopping is less important than for
the case A.  It is evident, because part of the oxygen degrees of
freedom is excluded from consideration.  Expressions for the matrix
elements of transitions to the excited states
$F_{ll'}^{xy}, \ D_{ll'}^{y}$ one can easily get from
(\ref{40}), (\ref{39}), and (\ref{10}).  They are given in Appendices
A and B.  A correction to the energy is not so small as for the case A.
(see Table III).  It is close to $13\%$, a correction to the first hopping
is close to $4.0\%$.  Thus, the $t$-$J$ model may be valid.  Inclusion of
hopping at
next neighbors requires consideration of corrections in the form
(\ref{27})  because they are not relatively small.
 They have a complicated structure.
One can see a discrepancy in our approach.  We assume that (U$_{d}
-\Delta )\ll$U$_{d}$,  i.e. $\Delta  \approx $U$_{d}$,  while the realistic
region for $\Delta $ (for which $J\simeq 126$ meV) is close to 3.6 eV.
Such situation is explained by the importance of the oxygen states for
the three-band model.

\subsection{ The special case of the complete model}\ \
Here we consider the complete three-band model (\ref{14}) without
any constraints on the Gilbert space.  Let us put (U$_{d}-2\Delta )
=-2\mu _{0}t_{p}$.  Such choice of the parameters leads to
accidental degeneracy of the energies of the $qq$ and $dd$
singlets.  The $q-d$ singlets sector of the Hamiltonian (\ref{14})
is written as
\begin{eqnarray}
\label{42}
H^{2}_{loc}&&=\sum_{l} \left\{ \left[ E_{\psi }X^{\psi \psi}_{l}+
E_{\varphi } X^{\varphi \varphi }_{l}
\right.\right.
\nonumber\\
&&\left.
- (\Delta +\mu_{0}t_{p}) X^{\chi \chi}_{l}
+2\sqrt{2}t\lambda_{0}(X^{\chi \psi}_{l} +X^{\chi\varphi}_{l}
 + H.c.)\right] + \dots,
\end{eqnarray}
where $E_{\psi }=E_{\varphi }=E_{0}=U_{d}-2\Delta =-2\mu _{0}t_{p}$.
Simplification in this case is as follows: linear combination of
$|\psi \rangle$ and $|\varphi \rangle$ (\ref{12}) singlets
\begin{eqnarray}
\label{43}
| c2 \rangle =(1/\sqrt{2})(| \psi \rangle - | \varphi \rangle)
\end{eqnarray}
does not hybridize with the $qd$ singlet $| \chi \rangle$ (\ref{12}).
After diagonalization of $H^{2}_{loc}$ we have:
\begin{eqnarray}
\label{44}
H^{2}_{loc}&&=\sum_{l} \left\{ E_{-}X^{c1c1}_{l}+ E_{0}X^{c2c2}_{l}
+E_{+} X^{c3c3}_{l}
\right.
\nonumber\\
&&+2\mu_{0}t_{p} X^{\tilde{\varphi}\tilde{\varphi}}_{l}
 +\tilde{E}_{-}\left( X^{\tilde{c}\tilde{c}}_{l}
+ \sum_{m =\pm1,0}X_{l}^{t1m\ t1m}\right)
\\
&&\left.
 +\tilde{E}_{+}\left( X^{\tilde{b}\tilde{b}}_{l}
+ \sum_{m =\pm1,0}X_{l}^{t2m\ t2m}\right)-
 (\Delta +\mu_{0}t_{p}) \sum_{m =\pm1,0} X_{l}^{\tau m\ \tau m}
 \right\}\ ,
\nonumber
\end{eqnarray}
where $E_{\pm} = -\Delta_{2}\pm R$, $E_{0}=-2\mu _{0}t_{p}$,
 $\Delta _{2}=U_{d}/4 + 2\mu_{0}t_{p}$,
 $\tilde{E}_{\pm} =-\tilde{\Delta} \pm R_{1}$,
 $\tilde{\Delta}=U_{d}/4$,
  $R_{1} =(\tilde{\Delta }^{2} +4t^{2}\lambda _{0}^{2})^{1/2}$, and
  $R =(\tilde{\Delta }^{2} +16t^{2}\lambda _{0}^{2})^{1/2}$.
Thus, the real hybridization parameter is $16t\lambda_{0}/U_d$ !
It means that the perturbation approach over $t/U_{d}$
\cite{Zh1}, \cite{Pa1} is unjustified for this system.  The
states $| \tilde{c}\rangle, |\tilde{b}\rangle,
| t1\rangle$ and $| t2\rangle$ are defined in Eq. (\ref{37})
\begin{eqnarray}
\label{45}
&&|c1\rangle= - a(|\varphi \rangle + |\psi \rangle)/\sqrt{2}+
b |\chi \rangle \ ,\
|c3\rangle= b(|\varphi \rangle + |\psi \rangle)/\sqrt{2} +
a |\chi \rangle \ ,
\nonumber\\
&&b=((R+\tilde{\Delta})/2R)^{1/2} \ ,\
a=((R-\tilde{\Delta})/2R)^{1/2} \ ,
\end{eqnarray}
Since in this case $\Delta =$U$_{d}/2 +\mu _{0}t_{p}$, one can determine
an adequate value of U$_{d}$ at fixed $t,t_{p}$ and $J$.  In the case
being considered, the experimental value of $J$ is achieved at U$_{d}=8.13$
 eV and $\Delta  =5.1$ eV.
Expressions for the parameters of the $t$-$t'$-$J$ model are
\begin{eqnarray}
\label{46}
E_{c1}&& \equiv E_{-}=-(U_{d}/4+2\mu_{0}t_{p}) -R,
\nonumber\\
t_{ll'}&&=2t\lambda _{ll'}(UV+ b a)
+ t_{p}\mu _{ll'} (aV +bU)^{2}/2  \ .
\end{eqnarray}
The local energies and hopping parameters for
this special case are given in Table IV.

The matrix $F_{ll'}^{xy}$ and $D_{ll'}^{y}$ elements which are required
for the derivation of corrections to
the energy $E_{-}$, the hopping constants $t_{ll'}$ and the superexchange
constant $J$  are given in Appendices A, B.
As in the first limiting case, O-O hopping plays an important
role for the magnitude of effective hopping of the lowest $c1$-singlet.
Correction to the energy is close to
$4.6\%$, correction to the first hopping constant is close to $2.5\%$.
Thus, the $t$-$J$ model is valid with the same precision.  Absolute
magnitude of the second hopping integral (hopping at the next-nearest
neighbors) is small due to the partial compensation of the amplitudes of
Cu-O and O-O hopping (see Eq. (\ref{46}) and Table IV).
As a result, correction to them is rather large $\approx 78\%$, and
the $t$-$t'$-$J$ model is not valid.  Relative magnitude of terms
with transition at the next neighbors (farther than nearest)  is of
the order of $10\%$.  This is also the parameter of accuracy of the
$t$-$J$ model as the low-energy limit of the three-band model.

\section{The general case} \ \ \
  In this part of the work we consider reduction of the three-band
model to the effective $t$-$t'$-$J$ model in the general region of
parameters when the charge transfer insulator is the ground state of
undoped system.  The general case differs from the special case
considered above only by the absence of degeneracy of the local $|
dd\rangle$ and $| qq\rangle$  two-hole states.  The local Hamiltonian is
defined by Eq.(\ref{42}) with  $E_{\psi }=$U$_{d}-2\Delta $ and
$E_{\varphi} = -2\mu t_{p}$.  We stress that $E_{\psi } \neq
E_{\varphi }$ in this case.  The expression for the diagonalized
Hamiltonian $H_{loc}^{2}$ coincides completely with Eq.(\ref{44}) but
the eigenstates $| ci\rangle$ have the more general form
\begin{eqnarray}
\label{47}
| ci \rangle =U_{i}| \psi \rangle + V_{i} | \varphi \rangle
 + W_{i} | \chi \rangle,
\end{eqnarray}
where coefficients $U_{i}, V_{i}, W_{i}$ are determined from
the solution of a system of three linear equations.  The   energies
$E_{-},E_{0},E_{+} $ are roots of the corresponding cubic equation.
Effective hopping parameters are
\begin{eqnarray}
\label{48}
t_{ll'}&&=2t\lambda _{ll'}(\sqrt{2}U_{1}U - W_{1}V) (\sqrt{2}V_{1}V
- W_{1}U)
+ t_{p}\mu _{ll'}  (\sqrt{2}V_{1}V - W_{1}U)^{2}/2 \ ,
\end{eqnarray}
Other matrix elements which are required for the derivation of the
constants $F_{ll'}^{x,y}$  (\ref{22}), (\ref{20}), (\ref{30}) and
 $D_{ll'}^{y}$ (\ref{24}), (\ref{34}), (\ref{35}) are presented in
Appendices A,B.

Parameters of the effective $t$-$t'$-$J$ model are calculated at the
following values of parameters of the three-band model:
$t=1.4$ eV, $t_{p}=0.7$ eV, U$_{d}=8$ eV and $\Delta =5.1$ eV
according to different band calculations \cite{Pi1,Fi1,Hy2} and
the detailed analysis of experiment \cite{Esk0}.

In Table V  the magnitudes of the hopping parameters at the first four
neighbors are given.  The important role of O-O hopping is
shown: for the first hopping parameter it is close to $30\%$, for others
is close to $50\%$.  Relative magnitudes of corrections
to the energy and first hopping are $3\%$ and $2.5\%$
respectively.  Thus, the $t$-$J$ model is valid with the same
precision.  Due to compensation of contribution of the $p-d$ and
$p-p$ hopping to the effective hopping integral at the second
neighbors (they have the opposite sign), the correction to it is
not small ($82\%$).  Thus, the model with transitions at the neighbors
farther than nearest must include the terms of the form (\ref{27}),
which depend on filling and the spin state of neighbor sites.

Applicability of a perturbation scheme in the realistic region of
parameters of the three-band model is provided by smallness of the
ratio of the effective hopping parameters between different local
states to the energy gap between them.  This ratio is of the order of
$10\%$ which gives the accuracy of the $t$-$J$ model for the CuO$_{2}$ plane.
Relative magnitude of the $t'$ -terms (hopping at the next neighbors)
in the $t$-$t'$-$J$ model is also of the order of $10\%$.
In Table V  the fundamental ratio $t_{1}/J =4.2$ for U$_{d}=8$ eV is
presented. It only weakly depends on U$_{d}$ and $\Delta$.  This
value is slightly larger than the generally accepted $t/J\approx 3$.

Different parameters of the three-band model are known with different
accuracy.  Notice that the value of the superexchange constant for
La$_{2}$CuO$_{4}$  is known  with high accuracy which imposes
restrictions on the values of the parameters of the three-band model.
Therefore one can try to determine the value of the least known
parameter $\Delta $ as a function of other parameters of the model.  At
fixed values of $t$ and $t_{p}$ presented above and $J=126$ meV  we get
$\Delta$ as a function of U$_{d}$ in the physically reasonable region:
$\Delta \leq$ U$_{d}\leq12$ eV  in  Fig. 1.  One can see that the reasonable
region of $\Delta $ lies between $4.75 $ eV  and $5.75$ eV  which is in
complete agreement with the band calculations of Sushkov and Flambaum
\cite{Fl1} but larger than the generally accepted value
$\Delta  = 2.75-3.75$ eV \cite{Esk0}.

\section{ Inclusion of V$_{pd}$ and U$_{p}$  terms} \ \ \
Here we consider the role of the intrasite repulsion at the oxygens
U$_{p}$ and the intersite repulsion V$_{pd}$.  We suppose that the
main effect is taken into account ( strong hybridization $d$- and
$q$-states), so that one can take the above named terms as the
perturbation.  Therefore, we do not consider the contribution of these
terms to the second-order corrections to the energies and the hopping
integrals of the form (\ref{25}), (\ref{27}). Our aim is to get the
renormalization of the first hopping integral and the charge-transfer
energy due to the Hamiltonian $\Delta H$ (\ref{1x}).
Using the representation (\ref{5}) for $q_{l}$ and $\tilde{q}_{l}$
operators one can easily get
\begin{eqnarray}
\label{50}
\Delta H_{pd}&&=V_{pd}\sum_{l,l_{1},l_{2},\alpha ,\beta }
n^{d}_{l\alpha }\left\{  f_{qq}({\bf l_{1}-l,l_{2}-l})
q^{+}_{l_{1},\beta }q_{l_{2},\beta }  \right.
\nonumber\\
&&+\ f_{\tilde{q}q}({\bf l_{1}-l,l_{2}-l})
[\tilde{q}^{+}_{l_{1},\beta }q_{l_{2},\beta }+H.c.] \\
&&\left. + f_{\tilde{q}\tilde{q}}({\bf l_{1}-l,l_{2}-l})
\tilde{q}^{+}_{l_{1},\beta }\tilde{q}_{l_{2},\beta } \right\}\ ,
\nonumber
\end{eqnarray}
Since we are interested in the renormalization of the lowest singlet
parameters, we can omit all terms with antisymmetrical oxygen
operators $\tilde{q}$.  Similar procedure for U$_{p}$ - term leads to
\begin{eqnarray}
\label{51}
\Delta H&&=V_{pd}\sum_{l,l_{1},l_{2},\alpha ,\beta }
f_{l,l_{1},l_{2}}n^{d}_{l\alpha }q^{+}_{l_{1},\beta }q_{l_{2},\beta }
\nonumber\\
&&+U_{p}\sum_{l,l_{1},l_{2},l_{3}}
h_{l,l_{1},l_{2},l_{3}}(q^{+}_{l,\uparrow  }q_{l_{1},\uparrow  } )
(q^{+}_{l_{2},\downarrow }q_{l_{3},\downarrow } ) \ ,
\end{eqnarray}
where $f_{l,l_{1},l_{2}}=f_{qq}({\bf l_{1}-l,l_{2}-l})$,
 $h_{l,l_{1},l_{2},l_{3}}=h({\bf l_{1}-l,l_{2}-l,l_{3}-l})$,
and
\begin{eqnarray}
\label{52}
f_{qq}({\bf l_{1}-l,l_{2}-l})&&= \sum_{{\bf k,k'}}f_{{\bf k,k'}}
 \exp(i{\bf k}({\bf l_{1}-l})-i{\bf k'}({\bf l_{2}-l}))
\nonumber\\
 h({\bf l_{1}-l,l_{2}-l,l_{3}-l})&&= \sum_{{\bf k,k',k''}}h_{\bf k,k',k''}
\\
 &&\exp(i{\bf k}({\bf l_{1}-l})-i{\bf k'}({\bf l_{2}-l})+i{\bf k''}
({\bf l_{3}-l}))
\nonumber
\end{eqnarray}
with
\begin{eqnarray}
\label{53}
f_{k,k'}&&=2\{\cos(k_{x}/2)\cos(k'_{x}/2)\cos((k_{x}-k'_{x})/2)+
(x\rightarrow y)\}(1+\gamma _{{\bf k}})^{-1/2}(1+\gamma _{{\bf k'}})^{-1/2}
\nonumber\\
h_{k,k',k''}&&=2\{\cos(k_{x}/2)\cos(k'_{x}/2)\cos(k''_{x}/2)
\cos((k_{x}+k'_{x}-k''_{x})/2)
\\
&&+(x\rightarrow y)\}
(1+\gamma _{{\bf k}})^{-1/2}(1+\gamma _{{\bf k'}})^{-1/2}
(1+\gamma _{{\bf k''}})^{-1/2}(1+\gamma _{{\bf k+k'-k''}})^{-1/2}\ .
\nonumber
\end{eqnarray}
Thus, we have the four-fermion terms with the complicated relation
between many states at different sites.  All later arguments are
based on a very rapid decrease of the constant  $f_{ll'}$ and
$h_{l,l_{1},l_{2},l_{3}} $ with $| {\bf l_{n}}-{\bf l}| $.  Our
calculation gives $f_{0} \equiv f(0,0)= 0.9180 $, $f_{1}=f(1,0)
=f(0,1) =0.1343$, and $S_{f} \equiv \sum _{l_{1}\neq l_{2}\neq l}
f_{l_{1},l_{2},l_{3}} =0.0092$;  $h_{0}\equiv h(0,0,0) =0.29$  and
$h_{1}\equiv  h(1,0,0) =h(0,1,0) =h(0,0,1) = 0.0096$.
Therefore, one can turn from (\ref{51}) to the effective Hamiltonian
\begin{eqnarray}
\label{54}
\Delta H_{loc}^{eff}&&=
V_{pd}\sum_{ll',\alpha ,\beta }f_{ll'}
n^{d}_{l\alpha }n^{q}_{l' \beta }
+U_{p}\sum_{ll'} h_{ll'}n^{q}_{l\uparrow}
n^{q}_{l'\downarrow}
\ ,
\nonumber\\
\Delta H_{hop}^{eff}&&=
V_{pd}f_{1}\sum_{\langle ll'\rangle,\alpha ,\beta }n^{d}_{l\alpha }\{
q^{+}_{l,\beta }q_{l',\beta } + H.c.\}
\\
&&+U_{p}h_{1}\sum_{\langle ll'\rangle}\left\{n^{q}_{l\uparrow }\{
q^{+}_{l,\downarrow }q_{l',\downarrow } + H.c.\}
+n^{q}_{l\downarrow }\{
q^{+}_{l,\uparrow }q_{l',\uparrow } + H.c.\} \right\} \ ,
\nonumber
\end{eqnarray}
where in $\Delta H_{loc}^{eff}$  $f_{ll'}\equiv f({\bf l'-l,l'-l})$.

Let us consider the system at unit filling and calculate the shift
of the energies of the local states due to $\Delta H_{loc}^{eff}$.
At unit filling there is the state
$|f\alpha \rangle =U| d\alpha\rangle - V| q\alpha \rangle$
at each site.  Hence, the shift of the
energy of $d$ - hole at the site $l$ will be
\begin{eqnarray}
\label{55}
\Delta \epsilon_{d}=V_{pd}V^{2}\sum_{l\neq l'} f_{ll'}=V_{pd}V^{2}(2-f_{0})
\end{eqnarray}
for the $q_{l}$ state
\begin{eqnarray}
\label{56}
\Delta \epsilon_{q}=V_{pd}U^{2}(2-f_{0})+U_{p}V^{2}(1/2-h_{0})
\end{eqnarray}
numbers 2 and 1/2 are the sums over all $l'$ for $f_{l',l}$
and $h_{l',l}$ respectively.
The coefficients $U$ and $V$ are defined in Eq.(\ref{16}) with
\begin{eqnarray}
\label{57}
R_{1}=(\tilde{\Delta}^{2} + 4t^{2}\lambda_{0}^{2})^{1/2} ,\ \ \
\tilde{\Delta}=(\epsilon_{q}-\epsilon_{d})/2\ ,
\end{eqnarray}
Thus, Eqs.(\ref{54}), (\ref{55}), (\ref{56}), and  (\ref{57}) are
the system of equations, which can be solved
numerically.   After solving we have new values $\tilde{U}$ and
$\tilde{V}$, and consider one hole over unit filling.  It is easy to
get the shift of the local energies of the additional hole.  For some
details see Appendix D.

The matrix elements of $\Delta H_{hop}^{eff}$ (\ref{53}) produce the
addition to the first hopping integral.
\begin{eqnarray}
\label{58}
\Delta t_{1}=(-f_{1}V_{pd}\tilde{U} W_{1}+\sqrt{2}h_{1}U_{p}\tilde{V}V_{1})
(W_{1}\tilde{U}-\sqrt{2}V_{1}\tilde{V})
\end{eqnarray}
The expressions for the matrix elements $\Delta D_{ll'}^{y}$ which are
essential for the constant J (\ref{33}), are given in Appendix B.

Different band calculations \cite{Pi1,Fi1,Hy2} give the consistent
magnitudes for V$_{pd}$ and U$_{p}$: V$_{pd} = 1.2$ eV, U$_{p}=4$ eV.
The calculations present no special problems.
Fig. 2 and Table VI present our results.  The lower curve in Fig. 2
demonstrates that for reasonable region of U$_{d}$  ($5-11$ eV), at fixed
V$_{pd}$, U$_{p}$, $t$ and $t_{p}$,   $\Delta  $ varies in the region
$2.5-3.5$ eV, which coincides very well with the band calculations
\cite{Hy2} and analysis of experiments \cite{Esk0}.

Different contributions to the first hopping parameter $t1$ are presented
in Table VI for U$_{d}=8.0$ eV  and $\Delta =3.0$ eV. Inclusion of
$\Delta H$ tends to decrease $t_{1}$. The ratio $t_{1} /J =3.38$,
which is close to the results of the other works \cite{Fi1,Hy1}.

\section{ Stability of the system} \ \ \
There are few conditions which determine stability of the system
under study. One can take an analogy from the classical
Hubbard system: the energy of two electrons at different sites
$(\{1_{l},1_{l'}\})$ is lower than the energy of two electrons at
the same site and an empty site $(\{2_{l},0_{l'}\})$ by the
energy of the Hubbard intrasite repulsion U$_{H}$.
Thus, the ground state of the system at unit filling is the state with
one electron per site which is a dielectric if U$_{H}$ is sufficiently
larger than band width $wzt$, where $z$ is the number of neighbors and
$w$ is the constant of the order of unity.

Since we have reduced our problem to the system of the Hubbard type
but with many states at each site, we can formulate the stability
conditions in terms of energies of these states.
The conditions will be
\begin{eqnarray}
\label{add1}
E^{2}+E^{0}-2E^{1}>0
\end{eqnarray}
and
\begin{eqnarray}
\label{add2}
E^{3}+E^{1}-2E^{2}>0\ ,
\end{eqnarray}
where the indexes $n=1,2,3$ denote the states with $n$ holes at site.
{}From this one can easily get the charge-transfer gap (an analogue of U$_{H}$)
as the difference between $(E^{2}+E^{0})$ and $2E^{1}$, where $E^{2}$ is
the energy of the lowest singlet and  $E^{1}$ is the energy of the
lowest one-hole state
\begin{eqnarray}
\label{add3}
\Delta E(2,0-1,1)=E_{-}+E^{0}-2E_{f}=3.2 \ \mbox{eV}>0
\end{eqnarray}
for the above mentioned parameters.

An interesting quantity is also the difference between
$(\{2_{l},2_{l'}\})$  and  $(\{3_{l},1_{l'}\})$ states
\begin{eqnarray}
\label{add4}
\Delta E(3,1-2,2)= E^{3}+E_{f}-2E_{-}=2.7 \ \ \mbox{eV} \ .
\end{eqnarray}
where $E^{3}$ is the lowest three-hole state (see Appendix C).
This energy difference $\Delta E(3,1-2,2)$ is also analogue of the energy
of the Hubbard repulsion U$_{H}$.
\section{  Conclusion} \ \ \
We conclude by summarizing our results.  We have studied the
low-energy properties of the CuO$_{2}$ plane near unit filling in the
framework of the three-band model.  We have considered the full
three-band model in the form that was put forward by Emery \cite{Em1}, in
the realistic region of the parameters, without any additional
assumptions about the smallness of some of them.

Thus, we have taken into account the direct O-O hopping, intrasite
repulsion at the oxygens U$_{p}$, and intersite repulsion V$_{pd}$,
which have not been considered earlier.  We have presented the consistent
approach to the mapping of the three-band model to it low-energy
limit.  Following the idea of Zhang and Rice \cite{Zh1}, we expressed
the primary Hamiltonian in terms of symmetrical and antisymmetrical
oxygen states.  Next, we turned to the terms of the local states
with certain number of the particles and transition between them.
The diagonalization of the local part of the Hamiltonian provided the
set of eigenstates.  The lowest of them is the singlet state with the
energy well below than others.
We have derived the low-energy Hamiltonian for the lowest singlet and its
transitions ($t$-$t'$-$J $ model), and have taken into account all upper
states by the special type of the unitary transformation.

Thus, we obtained the effective single-band Hamiltonian which
is essentially the $t$-$J$ model one. Transitions to the next
neighbors are not simple hopping due to the important role of the
correction, which has a complicated structure.

Our approach allows to establish the quantitative boundary of the
validity of the $t$-$J$ model as the low-energy limit of the three-band
model, to get the corrections to it in an explicit form, and to take
into account the transitions at the next neighbors.
It is evident, that one can determine the value of the charge- transfer
energy $\Delta $ from the well defined value of J at fixed other
parameters of the three-band model.  We have established that $\Delta $
varies in a narrow region of energies.

The lowest one-hole state for the undoped CuO$_{2}$ plane
is (\ref{16})
\begin{eqnarray}
\label{cx1}
|f\alpha\rangle=U |d\alpha\rangle - V |q\alpha\rangle \ \ \ \ \,
\alpha = \uparrow , \downarrow .
\end{eqnarray}
The lowest two-hole singlet state for the doped CuO$_{2}$
plane is (\ref{47})
\begin{eqnarray}
\label{cx2}
| c1 \rangle =U_{1}| d\uparrow d\downarrow \rangle +
V_{1} | q\uparrow q\downarrow \rangle  +
W_{1}(| d\uparrow q\downarrow  \rangle -
| d\downarrow q\uparrow  \rangle )/\sqrt{2}
\end{eqnarray}
where $|d\alpha\rangle$ is the state of a hole at the copper with projection
$\alpha$ ; $|q\alpha\rangle$ is the state of a hole  at the symmetrical
 oxygen
which represents the Wannier state formed from the four oxygen states around
the copper ion; $| d\uparrow d\downarrow \rangle  \ , \
| q\uparrow q\downarrow \rangle $ and
$(| d\uparrow q\downarrow  \rangle \ - \
| d\downarrow q\uparrow  \rangle )/\sqrt{2}$
are the two-hole singlet states.

The following set of parameters of the three band model
(\ref{1}) - (\ref{2}) seems the most acceptable at present:
$\Delta =\epsilon _{p}\ - \ \epsilon _{d}$ = 3 eV, $U_{d}$ = 8 eV,
$U_{p}$ = 4 eV, $V_{pd}$ = 1.2 eV, $t$ = 1.4 eV, $t_{p}$ = 0.7 eV.

For this set of parameters of the three band model we have the following
values of the coefficients $U,\ V,\ U_{1},\ V_{1},\ W_{1}$  (\ref{cx1}),
(\ref{cx2}) which determine the probability of location of holes at the
copper and the oxygen: $U$ = 0.85, $V$ = 0.52,
$U_{1}$ = -0.38, $V_{1}$ = -0.64, $W_{1}$ = 0.67.

The fundamental parameters of the t-J model t$_{1}$ and J
are : t$_{1}$ = 0.427 eV, J = 0.126 eV and t$_{1}$/J = 0.34.

The charge transfer gap $\Delta E$ = 3.2 eV and the effective Hubbard
repulsion of holes U$_{H}$ = 2.7 eV.

\newpage
{\Large \bf Acknowledgments} \\ \vskip 0.1cm
This work was supported, in part, by a Soros Foundation Grant awarded
one of us (V.I.B.) by the American Physical Society.
This work was  supported partly by the Soros-Akademgorodok
Foundation, by the Counsel on Superconductivity
of Russian Academy of Sciences, Grant No. 90214, and
by the Program  'Universities of Russia as the centers of
fundamental researches'.

\appendix
\section{Matrix elements for the singlet transitions}\ \
Here we present the explicit form for the matrix elements of the
transition of the additional hole from the lowest to excited states.
Thus,
\begin{eqnarray}
\label{A1}
F_{ll',\alpha }^{x\beta ,y}=\langle ly | \langle l'x\beta |
| H_{hop}| | cl' \rangle| f\alpha l\rangle\ ,
\end{eqnarray}
where $H_{hop}$ determined in eq.(\ref{10}).
A generator of the $S-W$ transformation is
\begin{eqnarray}
\label{A1a}
S&&=-\sum_{ll'\alpha \beta}\sum_{y}
\left.' \right.
\frac{F^{f\beta,y}_{ll',\alpha}}{E_{y}-E_{-}}
\left\{
X^{y,f\alpha}_{l}X^{f\beta,c}_{l'} - H.c. \right\}
\nonumber\\
&&-\sum_{ll'\alpha\beta ,y}
 \frac{F^{g\beta,y}_{ll',\alpha}}{E_{y}-E_{-}+\epsilon_{g}- \epsilon_{f}}
\left\{
X^{y,f\alpha}_{l}X^{g\beta,c}_{l'}- H.c.\right\} \ .
\end{eqnarray}
As was noted, the spin structure of the matrix elements can be
separated out (see Eqs. (\ref{29}), (\ref{30})).  So, for the first limiting
case (Sec. V.A) the matrix elements are
\begin{eqnarray}
\label{A2}
F_{ll'}^{f,b}&&=2t\lambda _{ll'}V (Uab-(b^{2}-a^{2})V/\sqrt{2})
- t_{p}\mu _{ll'}
(Vb\sqrt{2} - U a)(Ub+Va\sqrt{2})/2  \ ,
\nonumber\\
F_{ll'}^{f,\tau }&&=2t\lambda _{ll'}V ^{2}a
- t_{p}\mu _{ll'}U (Ub+Va\sqrt{2})/\sqrt{2}  \ ,
\nonumber\\
F_{ll'}^{f,\tilde{c}}&&= t_{p}\nu _{ll'}(Ub+Va\sqrt{2})/2  \ ,
\nonumber\\
F_{ll'}^{f,t1}&&=- t_{p}\nu _{ll'}(Ub+Va\sqrt{2})/\sqrt{2}  \ ,
\nonumber\\
F_{ll'}^{f,\tilde{b}}&&=F_{ll'}^{f,\tilde{\varphi }}=F_{ll'}^{f,t2}
\equiv 0 \ ,
\\
F_{ll'}^{g,c}&&=2t\lambda _{ll'} (b(U^{2}-V^{2})/2+aUV\sqrt{2})
- t_{p}\mu _{ll'}
(Vb-\sqrt{2}U a)(Ub+Va\sqrt{2})/2  \ ,
\nonumber\\
F_{ll'}^{g,b}&&=t\lambda _{ll'} (ab(U^{2}-V^{2})-UV(b^{2}-a^{2})\sqrt{2})
- t_{p}\mu _{ll'}
(Ua-Vb\sqrt{2})(Vb-Ua\sqrt{2})/2  \ ,
\nonumber\\
F_{ll'}^{g,\tau }&&=-\sqrt{2}t\lambda _{ll'}V (b-UVa\sqrt{2})
+ t_{p}\mu _{ll'}
U(Vb-\sqrt{2}U a)/\sqrt{2}  \ ,
\nonumber\\
F_{ll'}^{g,\tilde{c}}&&=- t_{p}\nu _{ll'}(Vb-Ua\sqrt{2})/2  \ ,
\nonumber\\
F_{ll'}^{f,t1}&&=- t_{p}\nu _{ll'}(Vb-Ua\sqrt{2})/\sqrt{2}  \ ,
\nonumber\\
F_{ll'}^{g,\tilde{b}}&&=F_{ll'}^{g,\tilde{\varphi }}=F_{ll'}^{g,t2}
\equiv 0 \ .
\nonumber
\end{eqnarray}
For the second limiting case (Sec. V.B.) there are matrix elements
only with $x=f$
\begin{eqnarray}
\label{A3}
F_{ll'}^{f,b}&&=-\sqrt{2}t\lambda _{ll'}(b^{2}-a^{2})
+ t_{p}\mu _{ll'}ba/2  \ ,
\nonumber\\
F_{ll'}^{f,\tau }&&=2t\lambda _{ll'} a
+ t_{p}\mu _{ll'}b/\sqrt{2}  \ ,
\nonumber\\
F_{ll'}^{f,\tilde{\chi }}&&= t_{p}\nu _{ll'}b/2  \ ,
\\
F_{ll'}^{f,\tilde{\tau }}&&= t_{p}\nu _{ll'}b/\sqrt{2}  \ .
\nonumber
\end{eqnarray}
The matrix elements for the third case (Sec. V.C.)
\begin{eqnarray}
\label{A4}
F_{ll'}^{f,c2}&&=t\lambda _{ll'}b(U^{2}-V^{2})
- t_{p}\mu _{ll'}
V(Ub+Va)/2  \ ,
\nonumber\\
F_{ll'}^{f,c3}&&=t\lambda _{ll'}(b^{2}-a^{2})
+ t_{p}\mu _{ll'}
((b^{2}-a^{2})VU-(U^{2}-V^{2})ba)/2  \ ,
\nonumber\\
F_{ll'}^{f,\tau }&&=\sqrt{2}t\lambda _{ll'}(U^{2}-V ^{2})a
+ t_{p}\mu _{ll'}U (Ub+Va)/\sqrt{2}  \ ,
\nonumber\\
F_{ll'}^{f,\tilde{c}}&&=- t_{p}\nu _{ll'}(Ub+Va)/2  \ ,
\nonumber\\
F_{ll'}^{f,t1}&&=- t_{p}\nu _{ll'}(Ub+Va)/\sqrt{2}  \ ,
\nonumber\\
F_{ll'}^{f,\tilde{b}}&&=F_{ll'}^{f,\tilde{\varphi }}=F_{ll'}^{f,t2}
\equiv 0 \ ,
\\
F_{ll'}^{g,c1}&&=t\lambda _{ll'} (U^{2}-V^{2})
- t_{p}\mu _{ll'}
(Vb-U a)(Ub+Va)/2  \ ,
\nonumber\\
F_{ll'}^{g,c2}&&=t\lambda _{ll'} (2UVb-a)
- t_{p}\mu _{ll'}
(Vb-U a)V/2  \ ,
\nonumber\\
F_{ll'}^{g,c3}&&= t_{p}\mu _{ll'}
(Ua-Vb)^{2}/2  \ ,
\nonumber\\
F_{ll'}^{g,\tau }&&=\sqrt{2}t\lambda _{ll'}V (2UVa-b)
+ t_{p}\mu _{ll'}
U(Vb-U a)/\sqrt{2}  \ ,
\nonumber\\
F_{ll'}^{g,\tilde{c}}&&=- t_{p}\nu _{ll'}(Vb-Ua)/2  \ ,
\nonumber\\
F_{ll'}^{f,t1}&&=- t_{p}\nu _{ll'}(Vb-Ua)/\sqrt{2}  \ ,
\nonumber\\
F_{ll'}^{g,\tilde{b}}&&=F_{ll'}^{g,\tilde{\varphi }}=F_{ll'}^{g,t2}
\equiv 0 \ .
\nonumber
\end{eqnarray}
The matrix elements for the general case (Sec.VI.) are
\begin{eqnarray}
\label{A5}
F_{ll'}^{f,ci}&&=-t\lambda _{ll'}[(\sqrt{2}  U_{i}U-W_{i}V)
(\sqrt{2}  V_{1}V-W_{1}U)+(\sqrt{2}  V_{i}V-W_{i}U)
 (\sqrt{2}  V_{1}V-W_{1}U)]
\nonumber\\
&&- t_{p}\mu _{ll'}(\sqrt{2}  V_{i}V-W_{i}U) (\sqrt{2}  V_{1}V-W_{1}U)/2 \ ,
\nonumber\\
F_{ll'}^{g,ci}&&=-t\lambda _{ll'}[(\sqrt{2}  U_{1}V+W_{1}U)
(\sqrt{2}  V_{i}V-W_{i}U)-(\sqrt{2}  V_{1}U+W_{1}V)
 (\sqrt{2}  U_{i}U-W_{i}V)]
\nonumber\\
&&+ t_{p}\mu _{ll'}(\sqrt{2}  V_{1}U+W_{1}V) (\sqrt{2}  V_{1}V-W_{1}U)/2 \ ,
\end{eqnarray}
where $i=1,2,3$ is the number of the singlet from $q-d$ sector. Other
matrix elements are
\begin{eqnarray}
\label{A6}
F_{ll'}^{f,\tau }&&=2t\lambda _{ll'}(U_{1}V ^{2}-U_{1}U^{2})
- t_{p}\mu _{ll'}U (\sqrt{2}V_{1}V-W_{1}U)/\sqrt{2}  \ ,
\nonumber\\
F_{ll'}^{f,\tilde{c}}&&= t_{p}\nu _{ll'} (\sqrt{2}V_{1}V-W_{1}U)/2  \ ,
\nonumber\\
F_{ll'}^{f,t1}&&=- t_{p}\nu _{ll'}(\sqrt{2}V_{1}V-W_{1}U)/\sqrt{2}  \ ,
\nonumber\\
F_{ll'}^{f,\tilde{b}}&&=F_{ll'}^{f,\tilde{\varphi }}=F_{ll'}^{f,t2}
\equiv 0 \ ,
\\
F_{ll'}^{f,\tau }&&=-\sqrt{2}t\lambda _{ll'}(W_{1}+(U_{1}+U_{1})UV\sqrt{2})
+ t_{p}\mu _{ll'}U (\sqrt{2}V_{1}U+W_{1}V)/\sqrt{2}  \ ,
\nonumber\\
F_{ll'}^{g,\tilde{c}}&&=- t_{p}\nu _{ll'} (\sqrt{2}V_{1}U+W_{1}V)/2  \ ,
\nonumber\\
F_{ll'}^{g,\tilde{c}}&&=- t_{p}\nu _{ll'}
 (\sqrt{2}V_{1}U+W_{1}V)/\sqrt{2} \ ,
\nonumber\\
F_{ll'}^{g,\tilde{b}}&&=F_{ll'}^{g,\tilde{\varphi }}=F_{ll'}^{g,t2}
\equiv 0 \ .
\nonumber
\end{eqnarray}

\section{Matrix elements for the virtual transitions}\ \
For the case of unit filling we need to determine the matrix
elements of the virtual transition of a hole at neighboring
clusters.  Thus, we are interested in the following quantities
\begin{eqnarray}
\label{B1}
D_{ll',\alpha\beta}^{y}=\langle ly | \langle l'0|
| H_{hop}| | f\alpha l' \rangle| f\beta l\rangle \ .
\end{eqnarray}
The spin structure can be separated out (see Eqs.
(\ref{34}), (\ref{35})).
A generator of the $S-W$ transformation has the form
\begin{eqnarray}
\label{B1a}
S=-\sum_{ll'\alpha \beta,y}
\frac{D^{y}_{ll',\alpha\beta }}{E_{y}+E_{0}-2\epsilon _{f}}
\left\{
X^{y,f\alpha}_{l}X^{0,f\beta}_{l'} - H.c. \right\} \ .
\end{eqnarray}
For the first limiting case (Sec.V.A.)
\begin{eqnarray}
\label{B2}
D_{ll'}^{c}&&=\sqrt{2}t\lambda _{ll'} (b+aUV\sqrt{2})
+ t_{p}\mu _{ll'}
V(Ub+Va\sqrt{2})/\sqrt{2}  \ ,
\nonumber\\
D_{ll'}^{b}&&=\sqrt{2}t\lambda _{ll'} (a-bUV\sqrt{2})
+ t_{p}\mu _{ll'}
V(Ua-Vb\sqrt{2})/\sqrt{2}  \ ,
\nonumber\\
D_{ll'}^{\tau }&&=2t\lambda _{ll'}(U^{2}-V ^{2})
+ t_{p}\mu _{ll'}UV  \ ,
\nonumber\\
D_{ll'}^{\tilde{c}}&&= t_{p}\nu _{ll'}V/\sqrt{2}  \ ,
\\
D_{ll'}^{t1}&&= t_{p}\nu _{ll'}V  \ ,
\nonumber\\
D_{ll'}^{\tilde{b}}&&=D_{ll'}^{\tilde{\varphi }}=D_{ll'}^{t2}
\equiv 0 \ .
\nonumber
\end{eqnarray}
For the second limiting case (Sec.V.B.)
\begin{eqnarray}
\label{B3}
D_{ll'}^{c}&&=-\sqrt{2}t\lambda _{ll'} b \ ,
\nonumber\\
D_{ll'}^{b}&&=-\sqrt{2}t\lambda _{ll'} a \ ,
\\
D_{ll'}^{\tau }&&=-2t\lambda _{ll'} \ ,
\nonumber\\
D_{ll'}^{\tilde{\chi }}&&= D_{ll'}^{\tilde{\tau }}
\equiv 0 \ .
\nonumber
\end{eqnarray}
For the third limiting case (Sec.V.C.)
\begin{eqnarray}
\label{B4}
D_{ll'}^{c1}&&=\sqrt{2}t\lambda _{ll'} (b+2aUV)
+ t_{p}\mu _{ll'}
V(Ub+Va)/\sqrt{2}  \ ,
\nonumber\\
D_{ll'}^{c2}&&= t_{p}\mu _{ll'}V^{2}/\sqrt{2}  \ ,
\nonumber\\
D_{ll'}^{c3}&&=\sqrt{2}t\lambda _{ll'} (2bUV-a)
- t_{p}\mu _{ll'}
V(Ua-Vb)/\sqrt{2}  \ ,
\nonumber\\
D_{ll'}^{\tau }&&=2t\lambda _{ll'}(U^{2}-V ^{2})
+ t_{p}\mu _{ll'}UV  \ ,
\nonumber\\
D_{ll'}^{\tilde{c}}&&= t_{p}\nu _{ll'}V/\sqrt{2}  \ ,
\\
D_{ll'}^{t1}&&= t_{p}\nu _{ll'}V  \ ,
\nonumber\\
D_{ll'}^{\tilde{b}}&&=D_{ll'}^{\tilde{\varphi }}=D_{ll'}^{t2}
\equiv 0 \ .
\nonumber
\end{eqnarray}
For the general case (Sec. VI.) matrix elements are
\begin{eqnarray}
\label{B5}
D_{ll'}^{ci}&&=-\sqrt{2}t\lambda _{ll'}(\sqrt{2}UV(U_{i}+V_{i})-W_{i})
- t_{p}\mu _{ll'}(\sqrt{2}  V_{i}V-W_{i}U) V/\sqrt{2}  \ ,
\nonumber\\
D_{ll'}^{\tau }&&=2t\lambda _{ll'}(U^{2}-V ^{2})
+ t_{p}\mu _{ll'}UV  \ ,
\nonumber\\
D_{ll'}^{\tilde{c}}&&= t_{p}\nu _{ll'}V/\sqrt{2}  \ ,
\\
D_{ll'}^{t1}&&= t_{p}\nu _{ll'}V  \ ,
\nonumber\\
D_{ll'}^{\tilde{b}}&&=D_{ll'}^{\tilde{\varphi }}=D_{ll'}^{t2}
\equiv 0 \ .
\nonumber
\end{eqnarray}
The additions to the matrix elements from U$_{p}$ and V$_{pd}$ terms,
which are essential for calculation of $J$ (Sec.  VII)
\begin{eqnarray}
\label{B6}
\Delta D_{\langle ll'\rangle}^{c1}&&=V_{pd} f_{1}UVW_{1}
-U_{p}h_{1}V_{1}V^{2}  \ ,
\nonumber\\
\Delta D_{\langle ll'\rangle}^{\tau }&&=V_{pd} f_{1}UV \ .
\end{eqnarray}

\section{Three-hole states}\ \
There are twenty of the three-hole cluster states
\begin{eqnarray}
\label{c1}
|\Psi _{\alpha }^{q}\rangle &&\equiv q_{\alpha }^{+}
d_{\uparrow }^{+}d_{\downarrow }^{+}
| 0 \rangle \ ,\ \
|\Psi _{\alpha }^{\tilde{q}}\rangle \equiv \tilde{q}_{\alpha }^{+}
d_{\uparrow }^{+}d_{\downarrow }^{+}
| 0 \rangle \ ,\ \
\nonumber\\
|\Phi _{\alpha }^{d}\rangle &&\equiv d_{\alpha }^{+}
q_{\uparrow }^{+}q_{\downarrow }^{+}
| 0 \rangle \ ,\ \
|\Phi _{\alpha }^{\tilde{q}}\rangle \equiv \tilde{q}_{\alpha }^{+}
q_{\uparrow }^{+}q_{\downarrow }^{+}
| 0 \rangle \ ,\ \
\\
|\tilde{\Phi} _{\alpha }^{d}\rangle &&\equiv d_{\alpha }^{+}
\tilde{q}_{\uparrow }^{+}\tilde{q}_{\downarrow }^{+}
| 0 \rangle \ ,\ \
|\tilde{\Phi}_{\alpha }^{q}\rangle \equiv q_{\alpha }^{+}
\tilde{q}_{\uparrow }^{+}\tilde{q}_{\downarrow }^{+}
| 0 \rangle \ ,\ \
\nonumber\\
\mbox{and }
\nonumber\\
|\Lambda _{\alpha\beta \gamma }\rangle &&\equiv d_{\alpha }^{+}
q_{\beta }^{+}\tilde{q}_{\gamma }^{+}
| 0 \rangle \ .
\nonumber
\end{eqnarray}
THe hybridization term (\ref{10}) provides the transitions only between
\begin{eqnarray}
\label{C2}
&&|\tilde{\Phi}_{\alpha }^{d} \rangle \leftrightarrow
|\tilde{\Phi}_{\alpha }^{q} \rangle \ ,
\nonumber\\
&&|\Psi_{\alpha }^{q} \rangle \leftrightarrow
|\Phi_{\alpha }^{d} \rangle \ ,
\\
\mbox{and}
\nonumber\\
&&|\Psi_{\alpha }^{\tilde{q}}\rangle, |\Phi_{\alpha }^{\tilde{q}}\rangle
 \leftrightarrow  (2\sigma )|\Lambda _{\sigma -\sigma  \alpha} \rangle \ .
\nonumber
\end{eqnarray}
Four other states $|\Lambda _{\sigma \sigma  \alpha}\rangle $
are noninteracting and degenerating in energy
\begin{eqnarray}
\label{C3}
E^{3}_{\Lambda _{\sigma \sigma  \alpha}}=-\Delta
\end{eqnarray}
counted from $\epsilon _{p}=0$.

By diagonalization of the Hamiltonian (\ref{10})
one can easily get that the eigenfunctions
$|\Psi_{\alpha }^{\tilde{q}}\rangle, |\Phi_{\alpha }^{\tilde{q}}\rangle$ ,
and $|\Lambda _{\sigma -\sigma \alpha}\rangle$ in the three-holes sector
 are connected with the $q-d$ sector of the two-hole states (\ref{44}).
Three-hole eigenfunctions are
\begin{eqnarray}
\label{C4}
|G_{\alpha ,i}\rangle \equiv \tilde{q}_{\alpha }^{+}| ci\rangle \ ,
\end{eqnarray}
and their energies
\begin{eqnarray}
\label{C5}
E^{3}(G_{\alpha ,i})\equiv E^{2+\tilde{q}}_{i}=\mu _{0}t_{p}+E^{2}_{i} \ ,
\end{eqnarray}
where $E^{2}_{i}\equiv E_{\pm,0}$ are the energies of the local singlets
(\ref{47}). Diagonalization of the other sectors (\ref{C2}) provides the
eigenenergies
\begin{eqnarray}
\label{C6}
 \tilde{E}_{\pm}^{3} =-(\Delta+3\mu _{0}t_{p})/2 \pm
\left(((\Delta-\mu _{0}t_{p})/2)^{2}+4t^{2}\lambda _{0}^{2}\right)^{1/2} \ ,
\end{eqnarray}
for the ($\tilde{\Phi}_{\alpha }^{d},\tilde{\Phi}_{\alpha }^{q}$) sector
\begin{eqnarray}
\label{C7}
 E_{\pm}^{3} =-(U_{d}-3\Delta-3\mu _{0}t_{p})/2 \pm
\left(((U_{d}-\Delta+\mu _{0}t_{p})/2)^{2}
+4t^{2}\lambda _{0}^{2}\right)^{1/2} \ ,
\end{eqnarray}
and for the ($\Psi_{\alpha }^{q},\Phi_{\alpha }^{d}$) sector.

Since we are interested in the lowest three-hole state, we analyze the
spectrum of the energies (\ref{C5}), (\ref{C6}). For a wide region of
U$_{d}$ and $\Delta $
\begin{eqnarray}
\label{C8}
E^{2+\tilde{q}}_{1}<E^{3}_{-} \ ,
\end{eqnarray}
and all other states lie much higher.
Renormalization of these energies due to the U$_p$ and V$_{pd}$ terms
remains all our consideration to be correct.

For the above mentioned parameters U$_{d}=8$ eV, $\Delta=3$ eV,
U$_{p}=4$ eV, and V$_{pd}=1.2$ eV one can get
\begin{eqnarray}
\label{C9}
E^{2+\tilde{q}}_{1}&&= -0.5  \ \ \mbox{eV}   \ ,
\nonumber\\
E^{3}_{-}&&= -3.73  \ \ \mbox{eV} \ .
\end{eqnarray}
Thus, the Hubbard repulsion U$_{H}$ is
\begin{eqnarray}
\label{C10}
U_{H}\ =\ E^{3}+E^{1}-2E^{2}= -3.73 - 4.27 + 2\cdot 5.36 \simeq 2.7 \ \
\mbox{eV} \ .
\end{eqnarray}

\section{Shifts of energies of the two-hole states}\ \
Shifts of energies of the primary set of the two-hole states,
due to U$_{p}$ and V$_{pd}$ terms are
\begin{eqnarray}
\label{d1}
\Delta E_{\varphi }&&=2V_{pd} U^{2}(2-f_{0})
+U_{p}(h_{0}+V^{2}(1/2-h_{0}))  \ ,
\nonumber\\
\Delta E_{\psi }&&=2V_{pd} V^{2}(2-f_{0})\ ,
\nonumber\\
\Delta E_{\chi }&&=2V_{pd} +U_{p}V^{2}(1/2-h_{0}))  \ ,
\\
\Delta E_{\tau }&&=2V_{pd} +U_{p}V^{2}(1/2-h_{0}))  \ .
\nonumber
\end{eqnarray}

\newpage
\vskip 0.7 cm
\noindent
TABLE I.  \\ {\small Coefficients $\lambda ({\bf l}-{\bf l'})$,
 $\mu ({\bf l}-{\bf l'})$, and $\nu ({\bf l}-{\bf l'})$ as  functions of
 $({\bf l}-{\bf l'})=n{\bf x}+m{\bf y}$ }.
\vskip 0.5 cm
\begin{center}
\begin{tabular}{cccc}
\hline \\
 n,m  & $\lambda _{n,m}=\lambda _{m,n}$ & $\mu _{n,m}=\mu _{m,n}$ &
$\nu_{n,m}=-\nu _{m,n}$ \\  \\ \hline \\
 0 , 0  & 0.9581 & 1.4567 & 0.0  \\
 1 , 0  & 0.1401 & 0.5497 & 0.2678   \\
 1 , 1  & -0.0235 & 0.2483 & 0.0      \\
 2 , 0  & -0.0137 & -0.1245 & 0.0812    \\
 2 , 1  & 0.0069 & -0.0322 & 0.0609      \\
 2 , 2  & 0.0035 & 0.0231 & 0.0  \\ \\ \hline
\end{tabular}
\end{center}
\vskip 1. cm
\noindent
TABLE II. \\
{\small Energies of the local states, the first four hopping parameters,
the second-order corrections to them on a ferromagnetic background,
correction to the energy $E_{-}$, the superexchange constant $J$,
the ratio of $(t_{1}/J)$ for the case $\Delta \ \ll $ U$_{d}$
at $\Delta =3.66$ eV.
\vskip 0.1 cm
\noindent
\begin{tabular}{ccccc}
\hline \hline \\
  one-hole local  &   E$_{f}=-5.33 $   &   E$_{g}=-0.65 $
&   E$_{\tilde q}=1.02$    \\
   energies, (eV)   &   \\
    two-hole local   &   E$_{-}=-7.38$   &   E$_{+}=6.57 $
&   E$_{\tau}=-4.68 $   \\
   energies, (eV)   &     $\tilde{E}_{-}=-4.31$    &
 $\tilde{E}_{+}=1.67$ &    E$_{\tilde \varphi }=2.04 $ \\ \\
\hline \\
 neighbor & direct hopping & contribution of & contribution of
& second-order  \\
 number,  & parameters, & direct O-O hopping, & Cu-O hopping, &
 corrections,  \\
 n & $t_{n}$, (eV) & $t_{n}^{pp}$, (eV) & $t_{n}^{pd}$, (eV) &
$\delta t_{n}$, ($\%$) \\ \\ \hline \\
 1  & 0.43 & 0.24 & 0.19 & 3.1 \\
 2  & 0.078 & 0.11 & -0.032 & 9.6  \\
 3  & -0.074 & -0.055 & -0.019 & 14.4 \\
 4  & 0.005 & -0.014 & 0.009 & 110.  \\
 \\ \hline \\
  correction to & & superexchange &  & ratio   \\
  the energy E$_{-}$, ($\%$) & & constant $J$, (meV) & &
 $|t_{1}/J|$  \\ \\ \hline \\
    4.1 & & 126.1 & & 3.4 \\ \\ \hline
\end{tabular}                                     }
\\
\newpage
\noindent
TABLE III.    \\
{\small Energies of the local states, the hopping parameters at the first
four neighbors, the second-order corrections to them
on a ferromagnetic background,
 correction to the energy E$_{-}$, the superexchange constant $J$,
and the ratio $t_{1}/J$ for the case U$_{d} - \Delta \ \ll $ U$_{d}$
at U$_{d}=8$ eV, $\Delta =3.65$ eV.
\vskip 0.1 cm
\noindent
\begin{tabular}{ccccc}
\hline \hline \\
  one-hole local  &   E$_{f}=-3.65 $  \\
   energies, (eV)   &   \\
 two-hole local & E$_{-}=-6.63$  &   E$_{+}=2.66$ & E$_{\tau}=-4.68 $   \\
   energies, (eV)   &     E$_{\tilde{\chi}}=-2.63$    &
 E$_{\tilde \tau}=-2.63 $ \\ \\
\hline \\
 neighbor & direct hopping & contribution of & contribution of
& second-order  \\
 number,  & parameters, & direct O-O hopping, & Cu-O hopping, &
 corrections,  \\
 n & $t_{n}$, (eV) & $t_{n}^{pp}$, (eV) & $t_{n}^{pd}$, (eV) &
$\delta t_{n}$, ($\%$) \\ \\ \hline \\
 1  & 0.38 & 0.15 & 0.23 & 4.6 \\
 2  & 0.030 & 0.068 & -0.038 & 313.  \\
 3  & -0.057 & -0.034 & -0.022 & 91. \\
 4  & 0.0022 & -0.0089 & 0.011 & 165.  \\
\\  \hline \\
  correction to & & superexchange &  & ratio   \\
  the energy E$_{-}$, ($\%$) & & constant $J$, (meV) & &
 $|t_{1}/J|$  \\ \\ \hline \\
   13.0 & & 126.6 & & 3.0 \\ \\ \hline
\end{tabular}                                     }
\\
\newpage
\noindent
TABLE IV. \\
{\small Energies of the local states, the hopping parameters at the first
four neighbors, the second-order corrections to them
on a ferromagnetic background,
 correction to the energy E$_{-}$, the superexchange constant $J$,
and the ratio $t_{1}/J$ for the special case
at U$_{d}=8.13$ eV.
\vskip 0.1 cm
\noindent
\begin{tabular}{ccccc}
\hline \hline \\
  one-hole local  &   E$_{f}=-6.42 $   &   E$_{g}=0.31 $
&   E$_{\tilde q}=1.02$    \\
   energies, (eV)   &   \\
    two-hole local   &   E$_{-}=-9.81$ & E$_{0}=-2.04 $ & E$_{+}=1.66 $
&   E$_{\tau}=-6.10 $   \\
   energies, (eV)   &     $\tilde{E}_{-}=-5.40$ &
 $\tilde{E}_{+}=1.33$ &    E$_{\tilde \varphi }=2.04 $ \\ \\
\hline \\
 neighbor & direct hopping & contribution of & contribution of
& second-order  \\
 number,  & parameters, & direct O-O hopping, & Cu-O hopping, &
 corrections,  \\
 n & $t_{n}$, (eV) & $t_{n}^{pp}$, (eV) & $t_{n}^{pd}$, (eV) &
$\delta t_{n}$, ($\%$) \\ \\ \hline \\
 1  & 0.53 & 0.19 & 0.34 & 2.5 \\
 2  & 0.028 & 0.085 & -0.057 & 79.5  \\
 3  & -0.076 & -0.043 & -0.033 & 22.9 \\
 4  & 0.0056 & -0.011 & 0.0166 & 25.2  \\
 \\ \hline \\
  correction to & & superexchange &  & ratio   \\
  the energy E$_{-}$, ($\%$) & & constant $J$, (meV) & &
 $|t_{1}/J|$  \\ \\ \hline \\
    4.66 & & 126.2 & & 4.18 \\ \\ \hline
\end{tabular}                                     }
  \\

\newpage
\noindent
TABLE V. \\
{\small Energies of the local states, the hopping parameters at the first
four neighbors, the second-order corrections to them
on a ferromagnetic background,
 correction to the energy E$_{-}$, the superexchange constant $J$,
and the ratio $t_{1}/J$ for the general case
at U$_{d}=8$ eV, $\Delta =5.1$ eV.
\vskip 0.1 cm
\noindent
\begin{tabular}{ccccc}
\hline \hline \\
  one-hole local  &   E$_{f}=-6.43 $   &   E$_{g}=0.31 $
&   E$_{\tilde q}=1.02$    \\
   energies, (eV)   &   \\
    two-hole local   &   E$_{-}=-9.85$ & E$_{0}=-2.12 $ & E$_{+}=1.61 $
&   E$_{\tau}=-6.12 $   \\
   energies, (eV)   &     $\tilde{E}_{-}=-5.41$ &
 $\tilde{E}_{+}=1.33$ &    E$_{\tilde \varphi }=2.04 $ \\ \\
\hline \\
 neighbor & direct hopping & contribution of & contribution of
& second-order  \\
 number,  & parameters, & direct O-O hopping, & Cu-O hopping, &
 corrections,  \\
 n & $t_{n}$, (eV) & $t_{n}^{pp}$, (eV) & $t_{n}^{pd}$, (eV) &
$\delta t_{n}$, ($\%$) \\ \\ \hline \\
 1  & 0.528 & 0.187 & 0.341 & 2.5 \\
 2  & 0.0274 & 0.0846 & -0.0572 & 82.2  \\
 3  & -0.0758 & -0.0424 & -0.0334 & 23.0 \\
 4  & 0.0057 & -0.011 & 0.0167 & 24.8  \\ \\
  \hline \\
  correction to & & superexchange &  & ratio   \\
  the energy E$_{-}$, ($\%$) & & constant $J$, (meV) & &
$|t_{1}/J|$  \\ \\ \hline \\
    3.02 & & 126.4 & & 4.18 \\ \\ \hline
\end{tabular}                                     }
\\
\newpage
\noindent
TABLE VI.                  \\
{\small Contributions of $p-d$ hopping, $p-p$ hopping, V$_{pd}$ and
U$_{p}$ terms to the effective hopping parameter of the lowest
singlet at the nearest neighbors, the superexchange constant $J$ and
the ratio $t/J$ at U$_{d}=8$ eV,  U$_{p}=4$ eV, V$_{pd}=1.2$ eV,
 $\Delta =3.0$ eV.
\vskip 0.1 cm
\noindent
\begin{tabular}{ccccc}
\hline \hline \\
  effective hopping &  $p-d$ & $p-p$ & V$_{pd}$ & U$_{p}$ \\
  parameter,
 & contribution,  & contribution,  & contribution,  & contribution,  \\
  $t_{1}$, (eV) &  (eV) & (eV) &  (eV) &  (eV)  \\ \\ \hline \\
 0.427 & 0.351 & 0.202 & -0.113 & -0.013  \\ \\  \hline \\
 &  superexchange &  & ratio &  \\
 & constant $J$, (meV) & & $|t_{1}/J|$ & \\ \\ \hline \\
  & 126.26  & & 3.38 &  \\ \\ \hline
\end{tabular}                                     }

\newpage
\vspace{2.cm}
\begin{thebibliography}{99}

\bibitem{An} P.W.Anderson, Science 235, 1196 (1987)

\bibitem{Em1} V.J.Emery, Phys. Rev. Lett. 58, 2794 (1987).

\bibitem{Va} C.M.Varma, S.Schmitt-Rink and E.Abrahams, Solid State
Commun. 62, 681 (1987).

\bibitem{Lok}Yu.B. Gaididei  and V.M. Loktev  Phys. Stat.Sol.(b) 147,
307 (1988)

\bibitem{Zh1} F.C.Zhang and T.M.Rice, Phys. Rev. B 37,
3759 (1988).

\bibitem{Zh2} F.C.Zhang and T.M.Rice, Phys. Rev. B 41,
 7243 (1990).

\bibitem{Sh1} J.L.Shen and C.N.Ting, Phys.Rev.   B 41,
1969 (1990).

\bibitem{Pa1} H.B.Pang, T.Xiang, Z.B.Su and L.Yu, Phys.Rev.
 B 41, 7209 (1990).

\bibitem{Yu1} S.V.Lovtsov and V.Yu.Yshankhai, Physica  C 179,
159 (1991).

\bibitem{Be1} V.I.Belinicher and A.L.Chernyshev, Phys.Rev.
 B 47, 390 (1993).

\bibitem{Za1} J.G.Zaanen, G.A.Sawtzky and J.W.Allen,
Phys. Rev. Lett. 55, 418 (1985)

\bibitem{Fr1} D.M.Frenkel, R.J.Gooding, B.I.Shraiman, and
E.D. Siggia, Phys.Rev. B 41, 350 (1990).

\bibitem{Pi1} W.E.Pickett, Rev.Mod.Phys. 61, 433 (1989).

\bibitem{Fi1} J.Fink at al., Physica C 185-189, 45 (1991).

\bibitem{Fl1} V.V.Flambaum and O.P.Sushkov, Physica C 175, 347 (1991).

\bibitem{Ma1} S.Maekawa, Y.Ohita, and T.Tohyama, Physica C 185-189,
 168 (1991).

\bibitem{Bi1} G.L.Bir and G.E.Pikus, Symmetry and deformational effects
in Semiconductors, Wiley, New York, 1974.

\bibitem{Hy1} M.S.Hybertsen, E.B.Stechel, M.Schluter and D.R.Jennison,
Phys.Rev. B 41, 11068 (1990).

\bibitem{Em2} V.J.Emery and G.Reiter, Phys.Rev. B 38,
 4547 (1988).

\bibitem{ZB} Weiyi Zhang and K.Bennemann, Phys. Rev. B 45,
12487 (1992).

\bibitem{Hy2} M.S.Hybertsen, W.M.C.Foulkes and M.Schluter,
Phys.Rev. B 45, 10032 (1992).

\bibitem{Esk0} H. Eskes, L.H.Tjeng, and G.A.Sawatzky,
Phys.Rev. B 41, 288 (1990).

\end{thebibliography}
\end{document}